\documentclass[epj,nopacs]{svjour}

\usepackage{graphicx}
\usepackage{bm} 

\newcommand{\GeV}{\makebox{ GeV}}
\newcommand{\beq}{\begin{equation}}
\newcommand{\enq}{\end{equation}}
\newcommand{\beqa}{\begin{eqnarray}}
\newcommand{\beqast}{\begin{eqnarray*}}
\newcommand{\enqa}{\end{eqnarray}}
\newcommand{\enqast}{\end{eqnarray*}}
\newcommand{\nn}{\nonumber}

\newcommand{\rf}{\ref}

\newcommand{\ga}{\gamma}

\newcommand{\ep}{\epsilon}

\newcommand{\si}{\sigma}

\newcommand{\om}{\omega}

\def\GeV{\nobreak\,\mbox{GeV}}

\def\half{\textstyle \frac{1}{2}}
\begin{document}
  
\title {  Nonperturbative and Perturbative Aspects
of Photo- and Electroproduction of Vector Mesons}
   \author{H.G. Dosch \inst{1} \and E. Ferreira \inst{2}} 
 \institute{Institut f\"ur Theoretische Physik, Universit\"at Heidelberg,   
  Philosophenweg 16, D-6900 Heidelberg, Germany, 
  \email{H.G.Dosch@thphys.uni-heidelberg.de} 
\and 
  Instituto de F\'{\i}sica, Universidade Federal do Rio de
 Janeiro, P.O. Box, Rio de Janeiro 21945-970, RJ, Brazil, 
  \email{erasmo@if.ufrj.br} }

\abstract{ We discuss various aspects of vector meson
production, first analysing the interplay between perturbative and
nonperturbative aspects of the QCD calculation. Using a general
method adapted to incorporate both perturbative and nonpertubative
aspects, we show that nonperturbative effects are  important for
all experimentally available values of the photon virtuality
$Q^2$. We compare the huge amount of experimental information now
available with our theoretical results obtained using a specific
nonperturbative model without free parameters, showing that quite
simple features are able to explain the data. 
  \keywords{electroproduction, vector mesons, wave functions,
  nonperturbative QCD, stochastic vacuum model}  }

\titlerunning{vector meson electroproduction}
\authorrunning{H.G.Dosch and E.Ferreira}
  \maketitle

\section{Introduction}\label{intro}

Electroproduction of  vector mesons provides an interesting
laboratory for studying the interplay between perturbative and
nonperturbative QCD. Most emphasis has been put on the
perturbative side of the
calculations\cite{DL87,pqcd1,pqcd2,pqcd3,pqcd4,pqcd5,pqcd6,pqcd7,pqcd8,pqcd9}. 
There the production process is considered to
be mediated by two gluon exchange and the coupling of the gluons
to the hadron, generally a proton, is taken from a
phenomenologically determined gluon distribution as obtained from
deep inelastic scattering (DIS). The coupling of the exchanged
photons to the produced quarks is treated perturbatively. This is
justified if there is a truly hard scale and the produced quarks
stay close together.

The emphasis put on perturbation theory is understandable given its
justification in terms of basic elements of QCD. Nevertheless we think
it is useful and even necessary to scrutinize also the other side of
the medal, namely the nonperturbative aspects, and investigate
its consequences. To examine the roles and magnitudes of different
effects we have chosen an approach which starts
from general  expressions and investigate the limit where perturbation
theory holds. We derive and discuss particularly the deviations from
perturbative QCD induced by nonperturbative effects.

There are several important reasons for this   approach.

\begin{itemize}

\item Even if the produced vector mesons are heavy, the effect of
binding by a confining potential is  not negligible at presently
accessible values of the photon virtuality $Q^2$. The binding
effect influences very strongly the $Q^2$ dependence of the
production amplitude.

\item Even for high values of $Q^2$ the production of transversely
polarized vector mesons receives important contributions from
regions where the two produced quarks are widely separated.

\item In spite of a clear transition from the perturbative to the
 nonperturbative regime, there are nevertheless striking systematic
features in the production of light and heavy vector mesons, whose
understanding requires  the incorporation of nonperturbative methods.

\item The gluon distribution of the proton as obtained from DIS
allows only to calculate production amplitudes with zero momentum
transfer. This requires quite essential extrapolations in the
analysis of the experimental data. Since the  approach discussed
here is based on a space-time picture it takes skewing effects
automatically into account, and therefore  the dependence of the
production amplitude on  (moderate) values of  $t$, the momentum
transfer from the virtual photon to the proton, can be calculated.

\item In QCD the production of vector mesons is closely related to
purely hadronic scattering processes without requiring the use of
vector dominance models. In order to obtain a unified picture, a
handling of nonperturbative effects is clearly necessary.

\item Closely related to the previous item is the relation between
QCD and Regge theory. In order to investigate possible bridges
between the two approaches again nonperturbative methods must be
used.

\end{itemize}

The calculation of photo and electroproduction presented in this
paper uses a general method for high energy scattering based on
the functional integral approach to QCD and on the WKB method,
which is capable of incorporating both the perturbative and
nonperturbative aspects \cite{Nac91,EN05}. The nonperturbative
input is given by a special model of nonperturbative QCD, called
stochastic vacuum model \cite{Dos87,DS88}, that has been
successfully applied in many fields, from hadron spectroscopy to
high energy scattering. The very satisfactory results of the model
in purely nonperturbative regions gives a strong weight to our
determination of nonperturbative correction terms near the
perturbative regime. The energy dependence is based on the
two-pomeron model of Donnachie and Landshoff \cite{DL98}.

In previous papers we have investigated  photoproduction \cite{DF02}
and electroproduction \cite{DF03}   of J/$\psi$ vector mesons. We
have also used the same framework to investigate some general
features of photo and electroproduction of all S-wave vector mesons
\cite{EFVL} which arise from the structure of the overlaps of
photon and vector meson wave functions.

   In the  recent years many more data have been obtained in HERA
experiments, with higher  accuracy and statistics, and their
comparison with theoretical calculations provide opportunities to
understand and describe important general features of the dynamics
governing these processes. In particular, we may  learn in what
extent the experimentally observed features are contained in the
global nonperturbative aspects of the systems in the final and
initial states, such as their wave functions and the long range
correlation properties of the intervening forces. In this paper we
describe very successfully most of these recent data, using the
same framework that has been tested in several other cases,
without the introduction of any free parameter.

Our paper is organized as follows.

In Sec. 2 we describe the methods used in the theoretical
calculations of photo and electroproduction of vector mesons.
Since this has already been done on several occasions,  we only
give a short and  schematic description, and then focus the
attention on the comparison with the usual perturbative treatment.

In Sec. 3 we present the results of our calculations of photo and
electroproduction of all 1S-wave vector mesons and compare them
with the experimental data. We show that the predictions for all
observables, the absolute values as well as the dependence on the
photon virtuality $Q^2$, the momentum transfer $t$ and the energy
$W$  are very satisfactory.

on our results and present them in context.

The appendix gives  formulas and details  of the theoretical
calculations and presents some of their general properties.

\section{General formul\ae~ \\and the strictly perturbative limit}
\label{general}

\begin{figure}
\includegraphics*[ width=7cm] {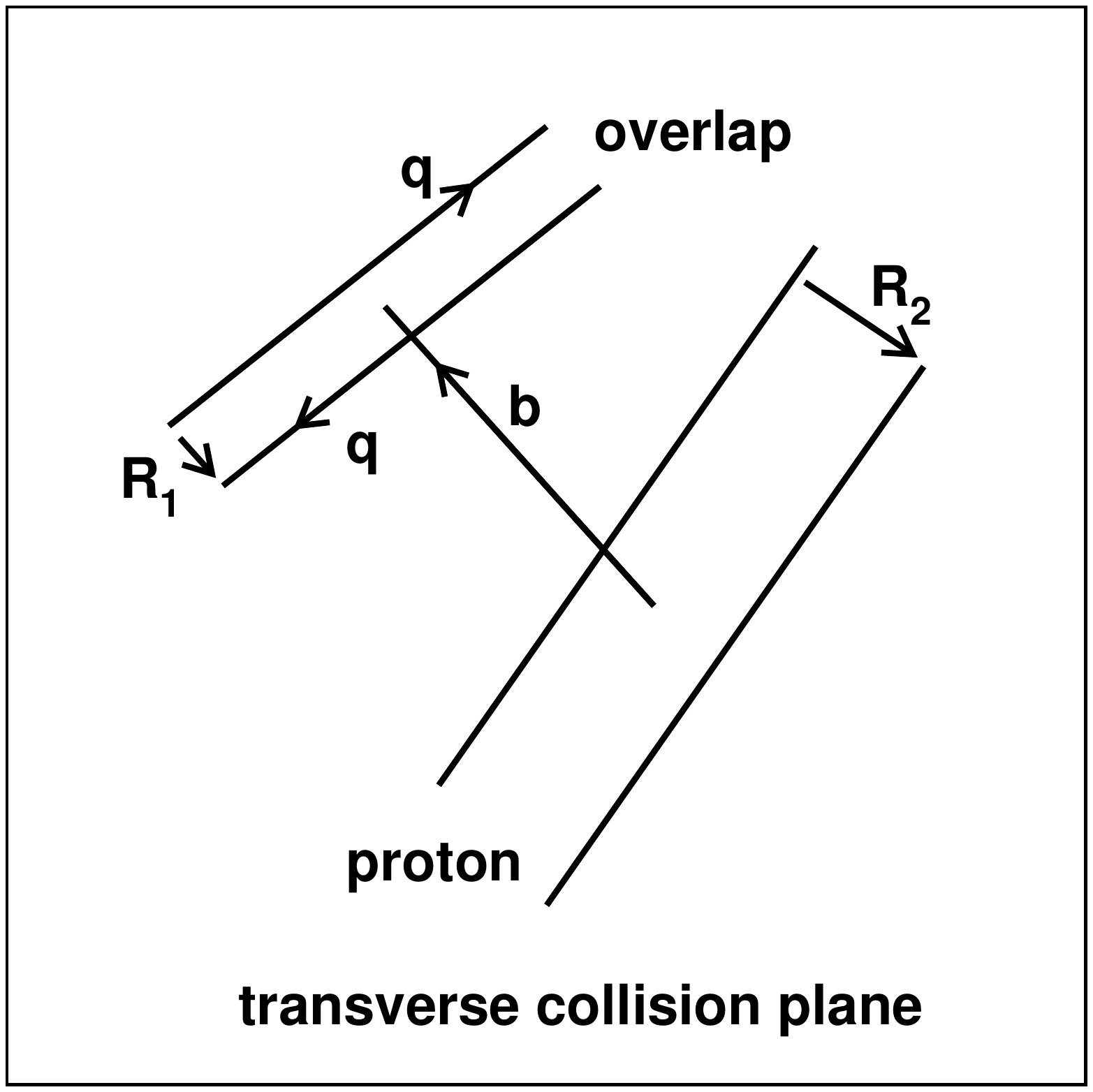} \hfill
\caption {  Loop-loop scattering.}
\label{loopsfig}
 \end{figure}

We start from a general approach to scattering based on functional
integrals and the WKB approximation 
\cite{Nac91,EN05}, which has
been adapted to hadron hadron scattering \cite{DK92,DFK94} and to
photo and electroproduction of vector mesons hadrons
\cite{DGKP97,DF02,DF03}. The basic features can be  seen in Fig.
\ref{loopsfig} that represents the loop-loop scattering amplitude
in real space time. The space time trajectory of the photon is
represented by a quark-antiquark loop, that of the proton by a
quark-diquark loop.
The transition to observable electroproduction
amplitudes of hadrons is achieved through a superposition of the
loop-loop amplitudes with the light cone wave functions of the hadrons
and the photon used as weights. This leads to  an
 electroproduction amplitude of the form
\beqa \lefteqn{ T_{\gamma^* p \to V p,\lambda}(W,t;Q^2)=(-2 i W^2)}      \\
&& \int d^2 R_1 dz_1
\psi_{V\lambda}(z_1,R_1)^*\psi_{\ga^* \lambda}(z_1,R_1,Q^2) 
J(\vec q,W,z_1,\vec R_1) ~ ,  \nonumber
\label{int} 
\enqa 
with 
\beqa \lefteqn{ J(\vec q,W,z_1,\vec R_1)=  } \\ 
&& \int  d^2 R_2 d^2 b \, e^{-i \vec q.\vec b}
~ |\psi_p(R_2)|^2 ~  S(b,W,z_1,\vec R_1,z_2=1/2,\vec R_2) ~ . \nonumber
\label{int2} \enqa 
Here $S(b,W,z_1,\vec R_1,1/2,\vec R_2)$ is the
scattering amplitude of two dipoles with separation vectors $\vec
R_1,~\vec R_2$, colliding with impact parameter vector  $\vec b$;
$\vec q$ is the momentum transfer vector of the reaction, with
\beq t = -\vec q\,^2 - m_p^2 (Q^2+M_V^2)/W^4 + O(W^{-6})\approx
-\vec q\,^2 ~ . \enq In these expressions  $Q^2=-p_{\gamma^*}^2$
is the photon virtuality, $W$ is the center of mass energy of the
$\gamma^*$-proton system, and $t=(p_{\rm V}-p_{\gamma^*})^2$ is
the invariant momentum transfer from  the virtual photon to the
produced vector meson; $z_1$ is the longitudinal momentum fraction
of the quark in the virtual photon and in the vector meson, and we
call $\bar z_1=1-z_1$; $\psi(z,\vec R)$ represent the light cone
wave functions.

The differential cross section is given by
\beq
\frac{d \si}{d|t|} = \frac{1}{16 \pi W^4} |T|^2 ~ .
\enq
For the special case of forward production ($\vec q=0$) this
approach reduces to the dipole model for electroproduction \cite{KNNZ94}.

Details of the evaluation of the loop-loop amplitude in the stochastic
vacuum model can be found in previous publications \cite{DF02,DGKP97}.

\subsection{The perturbative limits\\ and nonperturbative corrections}
\label{perturbative}

For the photon wave functions $\psi_{\ga^* \lambda}(z_1,R_1,Q^2)$
we use the well known lowest order
expressions, keeping the same notation used before \cite{DF02,DF03}.

If we concentrate on  high values
of virtuality $Q^2$ the two valence quarks stay close together
and we may use the dipole cross section of a small object of size $R_1$,
that is
\beq
J(\vec q=0,W,z_1,\vec R_1)= C(W,Q^2) \, R_1^2 ~.
\label{dipolcs}\enq

If we furthermore ignore the transverse extension of the vector
mesons, replacing
 \begin{equation} \psi_{V\lambda}(z_1,R_1) \to
\psi_{V\lambda}(z_1,0), \label{replace} \end{equation}
 we can
perform the $R_1$ integration in Eq. (\ref{int}) explicitly.

Later we introduce specific ans\"atze for the vector-meson wave functions
$\psi_{V\lambda}(z_1,R_1)$ and use the results of a specific model,
the stochastic vacuum model, for the evaluation of the loop-loop
scattering amplitude $J(\vec q,W,z_1,\vec R_1)$, but for the moment
we  stick to the general formulae and first study the corrections to
the strictly perturbative limit.

We first consider the simpler case of {\bf longitudinally
polarized} photons, where we obtain  with the replacement
(\ref{replace}) (for later discussion we keep the dependence on
the quark mass $m_f$ in our expressions):
 \beqa 
\lefteqn{T_{\gamma^* p\to V p,\lambda=0}^{\rm pert}=-(2iW^2)
16 C(W,Q^2)\,\hat e_f   }  \\
&& \,\frac{\sqrt{3\alpha}}{2 \pi}Q 
\int_0^1 d z_1\,\frac{8 \pi\, z_1^2 \bar z_1^2}{(z_1 \bar z_1
Q^2+ m_f^2)^2} \, \psi_{V 0}(z_1,0)~ . \nonumber
\label{pertl1} 
\enqa

The wave function at the origin $ \psi_{V 0}(z_1,0)$ is related to
the coupling of the vector meson to the electromagnetic current
$f_V$ by \beq f_V= \hat e_V \sqrt{3} \frac{1}{\sqrt{4 \pi}}
\int_0^1 d z_1\; 16 z_1 \bar z_1  \psi_{V 0}(z_1,0)~ .
\label{e_V}\enq

Here $\hat e_f$ is the quark charge, and $\hat e_V$ is the effective
charge in the meson, that is $\hat e_V= \textstyle{1}{\sqrt{2}}$ for
the $\rho$ and $\hat e_V= \textstyle{1}/{3\sqrt{2}}$ for the
$\omega$ meson , while for the $ \phi $, $\psi$ and $\Upsilon$ mesons
we have   $\hat e_V=\hat e_f$.

We introduce
 \beqa 
\lefteqn{
\eta_L(Q^2)=(Q^2/4 + m_f^2)^2 \, } \\ 
&& \int_0^1 d z_1\,
\frac{4\, z_1^2 \bar z_1^2\, \psi_{V 0}(z_1,0)} {(z_1 \bar z_1
Q^2+ m_f^2)^2} 
\Big/ \int_0^1 d z_1\; z_1 \bar z_1  \psi_{V0}(z_1,0) 
\label{etal}    \nonumber \enqa 
and then write   the perturbative expression as
 \beqa \lefteqn{T_{\gamma^* p\to V p,\lambda=0}^{\rm pert} = (-4
\pi i\, W^2)\, f_V\, } \\ 
&& \sqrt{\frac{\alpha}{\pi}} \frac{\hat e_f}{\hat
e_V} \eta_L(Q^2) \frac{Q}{(Q^2/4+m_f^2)^2} C(W,Q^2)~. \nonumber
\label{pertres}\enqa
Noting that the longitudinal wave function is suppressed at the
end points $z_1=0$, $z_1=1$, we see that in the limit
$Q^2\to \infty$ the mass $m_f$ can be neglected against
$z_1 \bar z_1 Q^2$ and the
expression $\eta_L$ becomes independent of $Q^2$
 \beq \eta_L \to
\frac{\int_0^1 d z_1\, \psi_{V 0}(z_1,0)}{ 4 \int_0^1 d z_1\; z_1
\bar z_1  \psi_{V 0}(z_1,0)} ~ .\label{asym}\enq
This is the correction factor due to the longitudinal extension of
the meson as obtained in \cite{pqcd2} (up to a factor 2 due to
different definitions). However, for practical purpose we keep
using   the expression (\ref{etal}) with finite quark masses. If
there were no binding effects, $z_1$ would be 1/2 and therefore
$\eta_L$ would approach 1 for $Q^2$ going to infinity.

Comparing Eq. (\ref{pertres}) with the perturbative expression for
the longitudinal electroproduction of vector mesons \cite{pqcd2}
  we obtain
\beq C(W,Q^2)= \frac{\pi^2}{3}x G(x,Q^2) \alpha_s(Q^2)
\label{relation}  ~ , \enq
where $x$ is the Bjorken variable $x=(Q^2+M_V^2)/W^2$.
This yields the well known relation
between the dipole cross section and the gluon density
\cite{BBFS93}.  After introducing our  specific  model
for the meson wave functions in subsection \ref{numerical}
we will present numerical values for $\eta_L(Q^2)$,
which represents the longitudinal momentum correction
to the pure perturbative calculation of the amplitude
for longitudinal photons.

Nonperturbative effects also lead to a finite extension of the
vector meson and to a deviation of the simple quadratic behaviour
of the ``dipole cross section'' $J(\vec q=0,W,z_1,R_1)$ in Eq.
(\ref{dipolcs}). We present numerical values for the rather large
corrections due to these effects   in subsection \ref{numerical}.

As it is well known, the treatment of the
{\bf transverse polarisation} is more delicate for at least
two reasons.
\begin{itemize}
\item
There is no strong suppression of the photon wave function at
the end points $z_1=0,~z_1=1$ and therefore the effective scale,
namely $z_1 \bar z_1\, Q^2$ might be quite low even for highly
virtual photons.
This makes among other things  a systematic $1/Q^2$ analysis
impossible since the factor corresponding to Eq. (\ref{asym})
diverges due to the singularities at the end points $z_1=0,\,1$.
\item
The meson wave function is supposed to be more complicated and
relativistic contributions are likely to be important even for
rather heavy mesons.
\end{itemize}

For simplicity we assume  that the wave function of the vector
meson has the same tensor structure as the transverse photon. Then
the  overlap function (\ref{int}) brings with the replacement
(\ref{replace}) the form 
\beqa &&T_{\gamma^* p\to Vp,\lambda= 1}^{\rm pert} =          
-2iW^2\,C(W,Q^2)\,\hat e_f \,\frac{\sqrt{6\alpha}}{2 \pi} \,  \\
&& \int_0^1 d z_1\,\frac{4\,m_f^2\, \psi_{V
1}(z_1,0) +16 ~ \omega^2 (z_1^2+\bar z_1^2) \psi_{V 1}^{(1)} (z_1,0) } {(z_1
\bar z_1 Q^2+ m_f^2)^2} ~ ,  \nonumber  \enqa
where
\beq
\psi^{(1)}_{V1}(z_1,0)=\frac{-1}{2 \omega^2}\,\left(\frac{
\partial}{\partial (R_1^2)} \psi_{V1}\right)(z_1,0) ~ . \enq
The relation to the decay constant $f_V$ is here  given  by
\beqa && f_V=  \hat e_V \frac{\sqrt{6}}{M_V} \frac{1}{\sqrt{4 \pi}} \\
&& \int_0^1 d z_1\,\frac{m_f^2 ~ \psi_{V 1}(z_1,0) +2 \omega^2 \,
(z_1^2 +\bar z_1^2)\, \psi_{V 1}^{(1)} (z_1,0)} {z_1 \bar z_1} ~ . \nonumber 
\enqa
This leads to a  relativistic correction factor $\eta_T$ which
is of order O($\textstyle {\omega^2}/{m_f^2}$), given by
\beqa
\label{etat} 
&& \eta_T(Q^2)=(Q^2/4 + m_f^2)^2 \,   \\  
&& \int_0^1 d z_1\,
\frac{ 4 m_f^2 ~ \psi_{V 1}(z_1,0)+ 16~  \omega^2  (z_1^2+\bar z_1^2)
 \psi_{V1}^{(1)}(z_1,0)}
{(z_1 \bar z_1 Q^2+ m_f^2)^2}    \nn \\
&& \Big/ \int_0^1 d z_1\,\frac{4 m_f^2 \psi_{V 1}(z_1,0) +8
\omega^2 \, (z_1^2 +\bar z_1^2)\, \psi_{V 1}^{(1)} (z_1,0)} {4\,z_1
\bar z_1} ~ .  \nn \enqa
The final result for the  for transverse
polarisation  is written analogous to Eq. (\ref{pertres}), that is
\beqa
&& T_{\gamma^* p\to V p,\lambda=1}^{\rm pert} = (- 4 \pi i\, W^2)
\,f_V\,\sqrt{\frac{\alpha}{\pi}} \frac{\hat e_f}{\hat e_V}  \nonumber \\
&& \eta_{T}(Q^2) \frac{Q}{(Q^2/4+m_f^2)^2} C(W,Q^2) ~ .
\label{pertrest}  \enqa
Only in the absence of binding effects it would be
$\omega=0,~z_1=\half$,  and we would then have $\eta_{T}(Q^2) \to
1$ for $Q^2 \to \infty$, as it is in the case of longitudinal
polarization. Numerical values for the correction coefficients
$\eta_T , \eta_L $ evaluated in a specific model are given in
subsection \ref{numerical}. Strong deviations of $\eta_L$ and
$\eta_T$ from unity indicate important deviations of the
amplitudes from the pure perturbative calculation, due to
longitudinal momentum in the photon-meson overlap and to
relativistic effects.

Important additional  corrections arise from the finite
extension of the vector mesons, and  they are also discussed
in subsection \ref{numerical}.

\subsection{The specific non-perturbative model}\label{model}

In order to obtain numerical information on the corrections
discussed in the two previous subsections, we need  to use
models both for the wave functions and for the interaction of
gluons and hadrons.

The photon enters through its usual perturbative wave function. In
model calculations~\cite{DGP98} it has been shown  that the
perturbative wave function can be used also for small values of
$Q^2+m_f^2$ if an appropriate constituent mass is introduced. For
light mesons, the masses have been determined by comparison with
the phenomenological two point function for the vector current.
The approach has been successfully applied in the theoretical
calculation of structure functions, Compton amplitudes and
photon-photon scattering.

The wave function has been adapted  from photons to vector mesons
including relativistic corrections motivated by the structure of
the vector current~\cite{DF03,DGKP97,DGP98,KDP99}. As in previous
papers \cite{DF02,DF03}, we use  two forms of meson wave
functions: the Bauer-Stech-Wirbel (BSW) \cite{BSW87}, and the
Brodsky and Lepage (BL) \cite{BL80} forms, which are detailed in
the Appendix. We use this type of wave function for all vector
mesons, with quark masses determined from a best fit to the vector
current (they are $m_u=m_d=0.2,~ m_s=0.3,~ m_c=1.25,~ m_b=4.2 ~$
GeV) .

Our nonperturbative treatment of the high energy process is based
on functional integration 
\cite{Nac91,EN05}
and on the stochastic
vacuum model \cite{Dos87,DS88}. The method has been described in
several occasions \cite{DFK94,DDLN02,DDSS02}, and we only quote
here a few of its characteristic features. The model is based on
the assumption that nonperturbative QCD can be approximated by a
Gaussian process in the coulour field strengths; the gluon field
correlator is therefore the quantity determining the full
dynamics. Its parameters are taken from lattice calculations
\cite{DGM99}. The model yields confinement in non-Abelian gauge
theories and leads to realistic quark-antiquark potentials for
heavy quarks\cite{DDSS02}. It can be used to determine the
loop-loop scattering amplitudes mentioned in the introduction and
visualized in Fig. \ref{loopsfig}.

For the energy dependence we have introduced in the model
\cite{DDR00,DD01,DD02,DF02} the two-pomeron scheme  of Donnachie and 
Landshoff \cite{DL98}. Small dipoles couple to the hard and large 
dipoles to the soft pomeron. The transition radius was determined 
through the investigation of the proton structure function.
Again we refer to the literature for more information and collect some
details and the relevant parameters  in the Appendix.

\subsection{Numerical results \\for the  nonperturbative corrections
\label{numerical}}

In order to exhibit the importance of the nonperturbative
contributions we display their effects explicitly in this
subsection. We hasten to add that for the final calculations as
given in Sec. \ref{results} we do not split our results into
perturbative and nonperturbative parts, but give directly the
full theoretical results.

The correction factors $\eta_L$ (\ref{etal}) and $\eta_T$
(\ref{etat}) are displayed in Fig. \ref{eta} as functions of $Q^2
+ 4 m_f^2$ for several vector mesons. The factor $\eta_L$ reflects
the effect of the distribution of the longitudinal momentum; we
notice that it remains of order 1, and its influence is rather
weak. In contrast, the correction in the transverse amplitude due
to the factor $\eta_T$, reflecting mainly relativistic corrections
to the transverse wave function, is very important, especially for
high values of $Q^2+m_f^2$. Its large values at high $Q^2$ for the
production of $\rho$-mesons indicate that the measurement of the
ratio of the longitudinal to the transverse production cross
sections tests mainly  the wave function. According to our
calculations, only the quantity $\eta_T$ for light mesons is very
sensitive to the specific choice  of the wave function.

\begin{figure}
\includegraphics*[ width=7cm] {etalbln.pdf} \hfill
\includegraphics*[ width=7cm] {etat2bln.pdf}
\caption  {Correction factors $\eta_L$ (\ref{etal}) and $\eta_T$,
(\ref{etat}) due to the longitudinal momentum distribution in the
longitudinal and transverse wave function of the mesons. The
factors are similar for the BL and BSW wave functions, except for
$\eta_T$ for light mesons. The differences in these cases are
illustrated  by the comparison of the long-dashed and short-dashed
curves corresponding  to the  BL (long-dashed)  and the BSW
(short-dashed)  wave functions for  the $\rho$-meson. \label{eta}}
\end{figure}

\begin{figure}
\includegraphics*[bb=130 360 350 520,width=7cm] {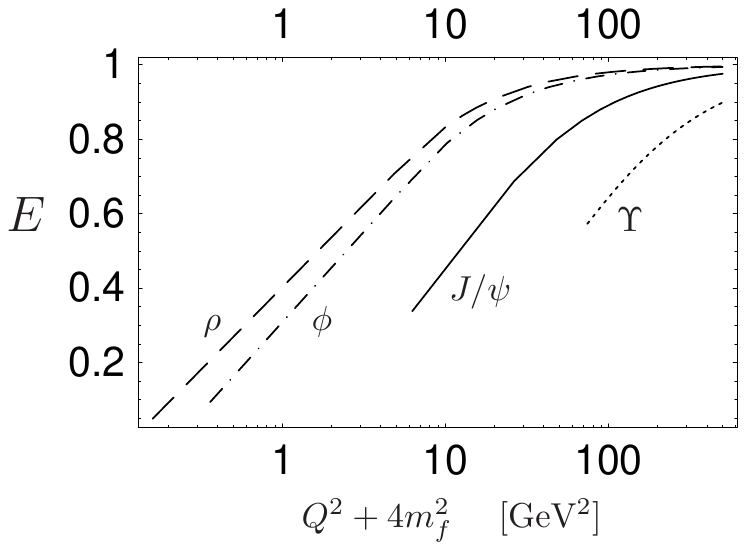}\hfill
\includegraphics*[bb=130 360 350 520,width=7cm] {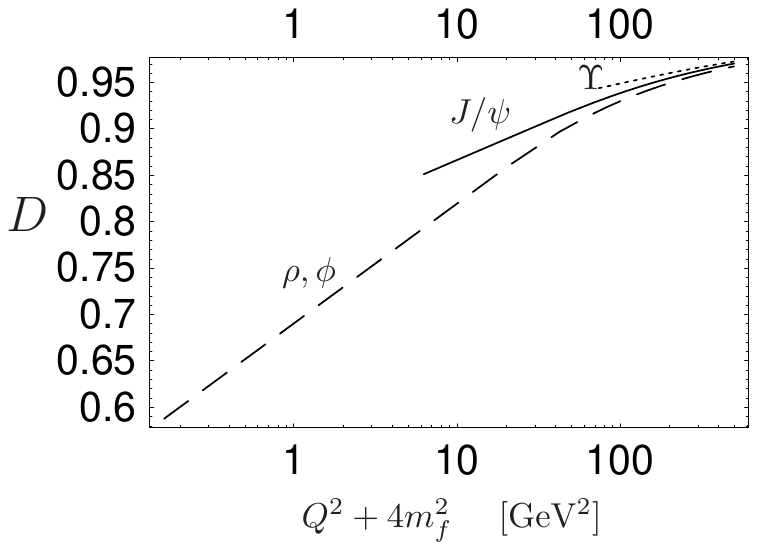}
\caption  {Correction factors in the evaluation of amplitudes for
electroproduction. The factor $E$ in the left is  due to the
finite extension of the meson wave function and the correction
factor $D$ in the right is due to nonperturbative corrections to
the simple $R_1^2$ dependence  of the dipole cross section. The
factors are very similar for longitudinal and transverse
polarisations. \label{extdip}}
\end{figure}

There are large effects due to  the finite extensions of the mesons.
 We denote by  $E$ the ratio of the full amplitudes (\ref{int}) to
the purely perturbative ones (\ref{pertres}),(\ref{pertrest}).The
ratio $E$ is displayed in Fig. \ref{extdip}, left, the results for
both polarisations are very similar.

The effects are by no means negligible, even for the heavy mesons.
The suppression due to the finite extension at low values of $Q^2$
leads, in the full range of presently available data,  to a much
weaker decrease in $Q^2$  than inferred from purely perturbative
expressions.

For large values of the quark-antiquark separation $R_1$, the
dipole cross section differs from the pure $R_1^2$-behaviour as
given by (\rf{dipolcs}),(\ref{relation}). We denote by  $D$ the
ratio of production amplitudes of the more realistic dipole cross
section  obtained with the stochastic vacuum model  divided by the
result obtained with the purely quadratic expression

light mesons at low values of $Q^2$. Also $D$ is  similar for the
longitudinal and transverse polarisation.

The importance of nonperturbative contributions to vector meson
photoproduction has been  stressed before \cite{DGS01},
using a different approach to nonperturbative corrections.

\section{\label{results} Theoretical results\\ and comparison with experiment}
 
A detailed description of the framework of our calculations can be
found in our previous papers \cite{DF02,DF03,DGKP97,EFVL}. The
basis for the determination of the loop-loop scattering amplitude
represented in Fig. \ref{loopsfig}, is the model of the stochastic
vacuum. The calculations are determined by two parameters, the
correlation length of the nonperturbative gluon fluctuations and
the strength of the gluon condensate, which can be  extracted from
lattice calculations~\cite{DGM99}.

All parameters have been determined from other sources, and they
are collected in Appendix A. Our calculations contain therefore no
adjustable parameter and all theoretical results are true
predictions.

In the following we present a comparison of our theoretical
results with the available data of elastic photo and
electroproduction of the S-wave vector mesons, namely
$J/\psi,\Upsilon,\rho,\omega,\phi$.

\subsection{Photo and electroproduction of the J/$\psi$ meson}

\begin{figure}[h]
 \vskip 2mmpdf
 \includegraphics[height=8cm,width=8cm]{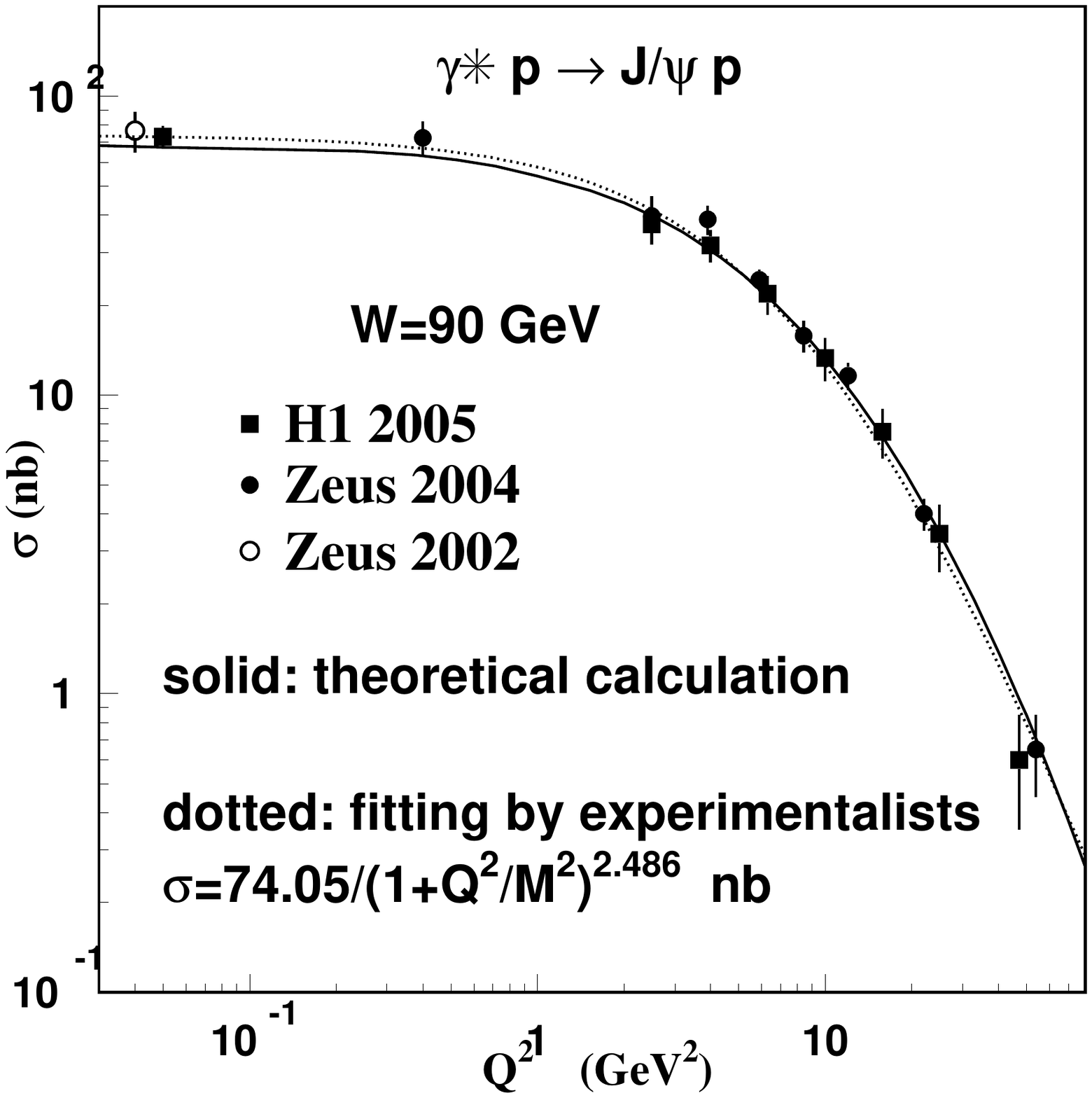}
 \caption{\label{psiq2} Integrated elastic cross
section of $J/\psi$ photo and electroproduction    at the energy
W=90 GeV as a function of $Q^2$. The data are from the Zeus
\cite{Zeus2002,Zeus2004} and H1 \cite{H12005} collaborations. The
solid line represents our theoretical calculation using the BL
wave function.The dotted curve is a pure fit to the experimental
points of the form given by Eq.(\ref{psiq2fit}).}
\end{figure}

In Fig. \ref{psiq2} we show the data for the integrated elastic cross
section $\sigma$ of $J/\psi$ electroproduction at the fixed energy
W=90 GeV as a function of the photon virtuality $Q^2$. The data are from
Zeus \cite{Zeus2002,Zeus2004} and H1 \cite{H12005} collaborations.
 The theoretical calculations (solid line) are made using the BL
(Brodsky-Lepage) form of the vector meson wave function; the
BSW-wave function yields very similar results. The agreement
between data and the theoretical calculations is remarkable.
The dotted line in the figure, which nearly coincides with our
theoretical calculation is a fit to the data proposed in the
experimental paper, with the usual form
\begin{equation}\label{psiq2fit}
\sigma=\frac{A}{(1+Q^2/M_V^2)^{n}} ~ .
\end{equation}

\begin{figure}[h]
 \vskip 2mm
 \includegraphics[height=8cm,width=8cm]{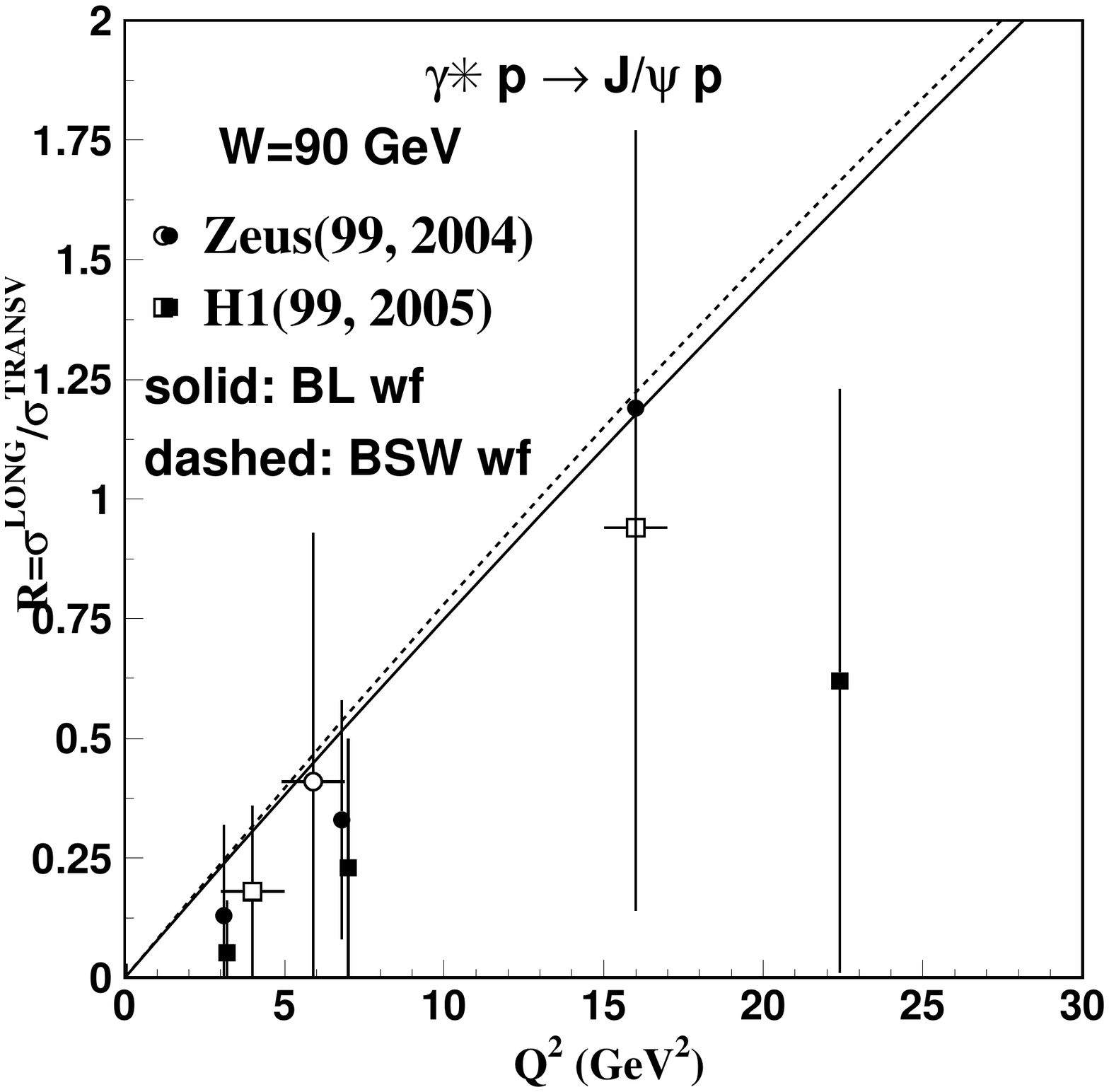}
 \caption{\label{psiratio} Ratio R of longitudinal and
transverse cross sections for $J/\psi$ electroproduction as
function of $Q^2$ for fixed energy W=90 GeV. Data from Zeus
\cite{Zeus99,Zeus2004} and H1 \cite{H199,H12005} collaborations at
HERA. The solid line and dashed lines show our theoretical
calculations respectively with the BL (solid) and the BSW (dashed)
wave functions. A good numerical representation for the BL result
is $R(Q^2) \approx 0.76~(Q^2/M^2)/(1+Q^2/M^2)^{0.09}$. }
\end{figure}

Fig. \ref{psiratio} shows the ratio $R$ of the longitudinal to the
transverse cross sections
\begin{equation}
\label{ratio}
 R=\frac{\sigma^L}{\sigma^T}  ~ ,
\end{equation}
again for $J/\psi$ elastic electroproduction at W=90 GeV. As the
ratio cancels influences of the specific dynamical model, this
quantity tests directly details of the overlaps of wave functions,
helping the study of their longitudinal and transverse structures
The data are from Zeus \cite{Zeus99,Zeus2004} and H1
\cite{H199,H12005} collaborations at HERA.  As seen in the figure,
the presently available data are not very accurate. The
theoretical calculations with the two kinds of wave function - BL
and BSW - give nearly the same results for
$\sigma=\sigma^L+\sigma^T$ and for $R=\sigma^L/\sigma^T$ .

\begin{figure}[h]
 \vskip 2mm
 \includegraphics[height=7cm,width=7cm]{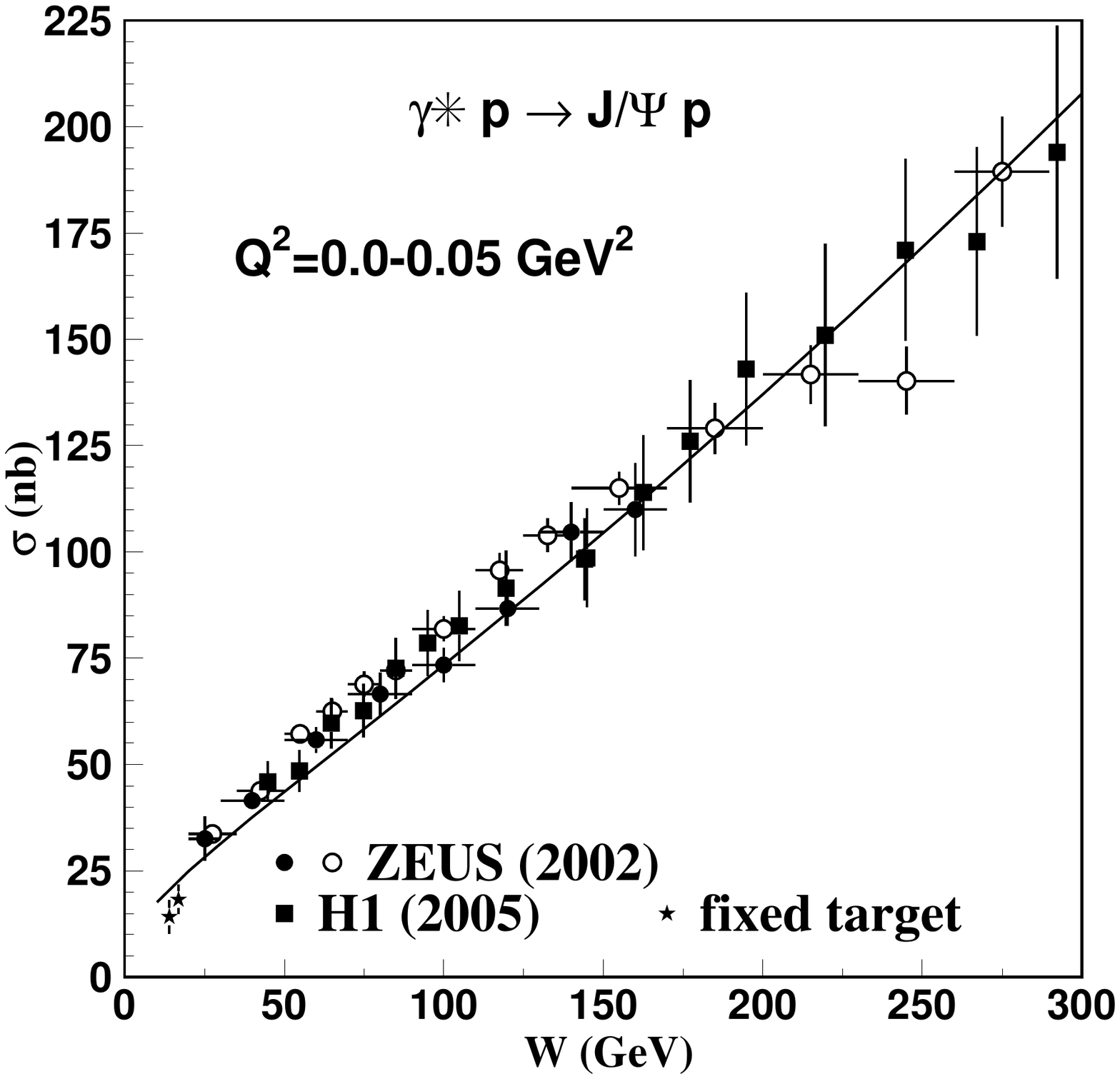}
 \includegraphics[height=7cm,width=7cm]{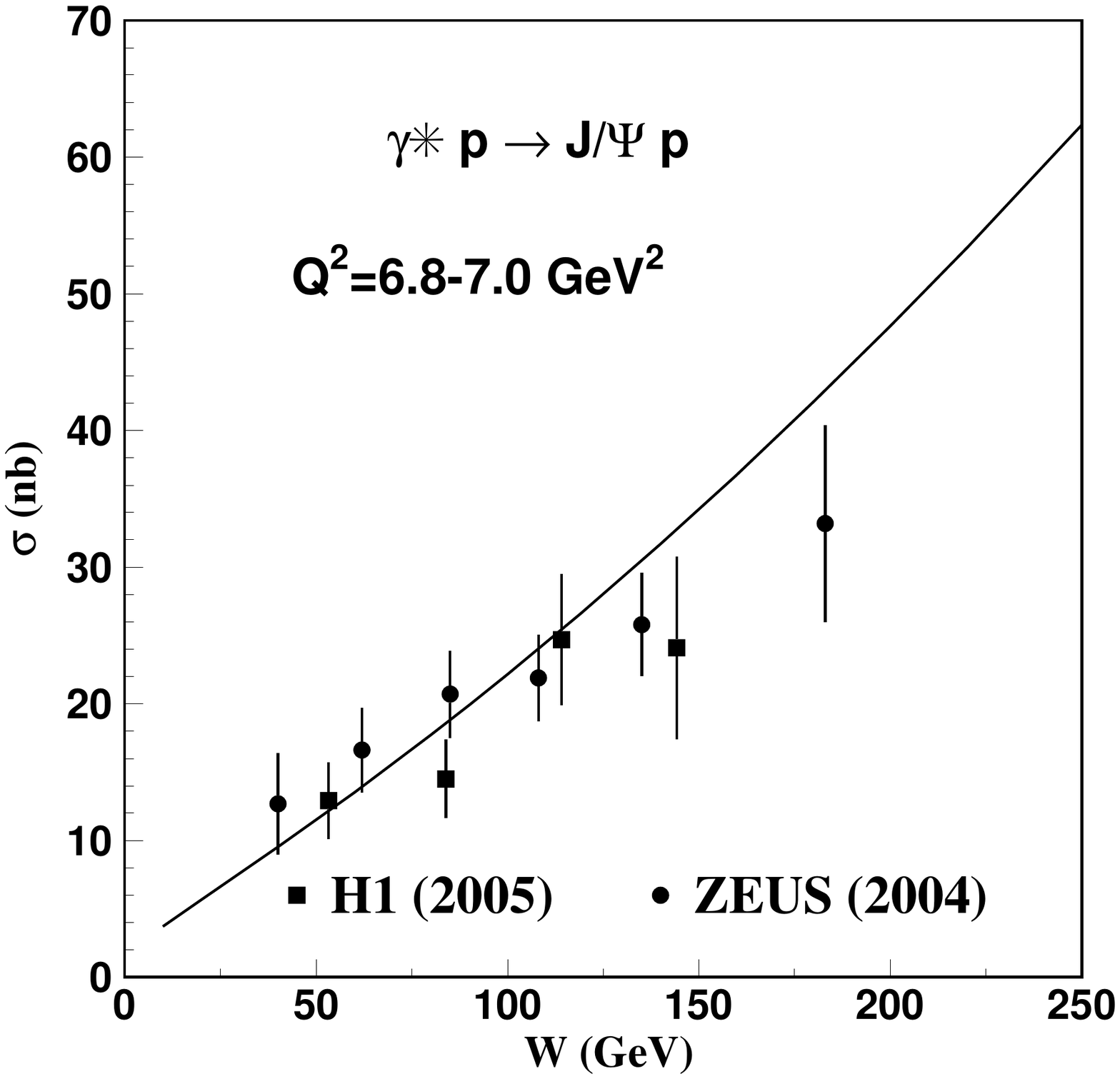}
 \caption{\label{psiW} Energy dependence of the integrated elastic
cross section for J/$\psi$ production from the HERA collaborations
\cite{Zeus2002,Zeus2004,H12005}, compared with our theoretical
calculations. The fixed target photoproduction data are from the
E-401 and E-516 experiments  \cite{E401,E516}. Electroproduction
is represented by the $Q^2$ range  6.8-7.0 GeV$^2$, for which
there are both Zeus and H1 data points. }
\end{figure}

Fig. \ref{psiW}   shows the comparison of our calculations for the
energy dependence of cross sections with the recent Zeus and H1
data \cite{Zeus2002,Zeus2004,H12005} and  older photoproduction
data from fixed target experiments \cite{E401,E516} at lower
energies. For illustration of our results for electroproduction we
show data and theoretical results in the  $Q^2$ range 6.8 - 7
GeV$^2$ for which there are both Zeus and H1 data.

\begin{figure}[h]
 \vskip 2mm
 \includegraphics[height=8cm,width=8cm]{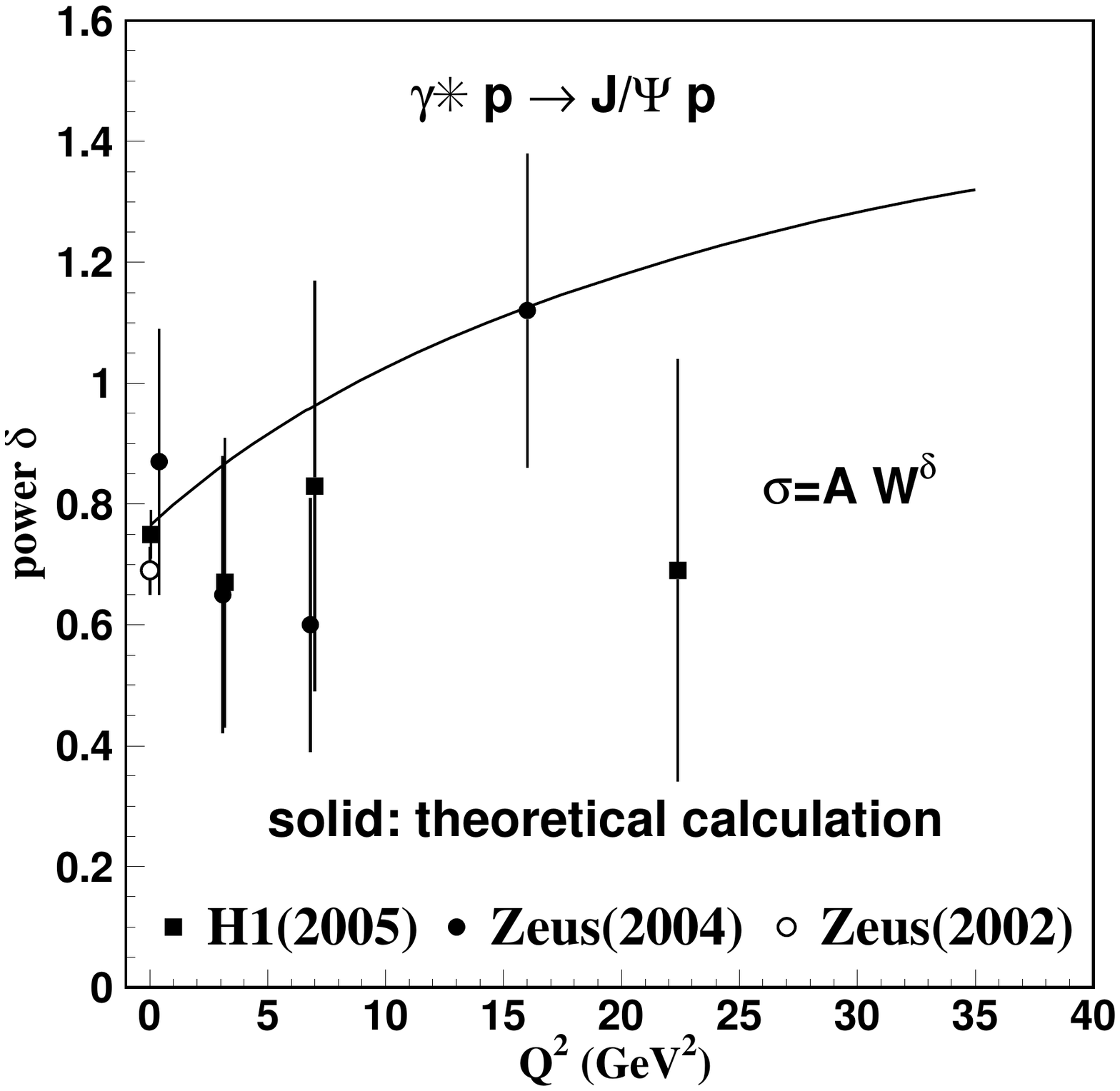}
 \caption{\label{delta_psi} $Q^2$ dependence of the $\delta$
parameter describing the energy dependence of the $J/\psi$
integrated elastic cross section, from  Zeus and H1 collaborations
\cite{Zeus2002,Zeus2004,H12005}, compared to our theoretical predictions.}
\end{figure}

Often the $W$  dependence of  experimental cross sections is fitted
through single powers in the form
\begin{equation}
\label{wpower}
 \sigma= {\rm Const.} \times  W^{\delta(Q^2)} ~ ,
\end{equation}
which may be useful in limited
$W$ ranges. Values of the parameter $\delta$ obtained at several values of
$Q^2$ in J/$\psi$ electroproduction are compared to our theoretical
predictions in Fig. \ref{delta_psi}, based on the two-pomeron scheme.

\begin{figure}[h]
 \vskip 2mm
 \includegraphics[height=7cm,width=7cm]{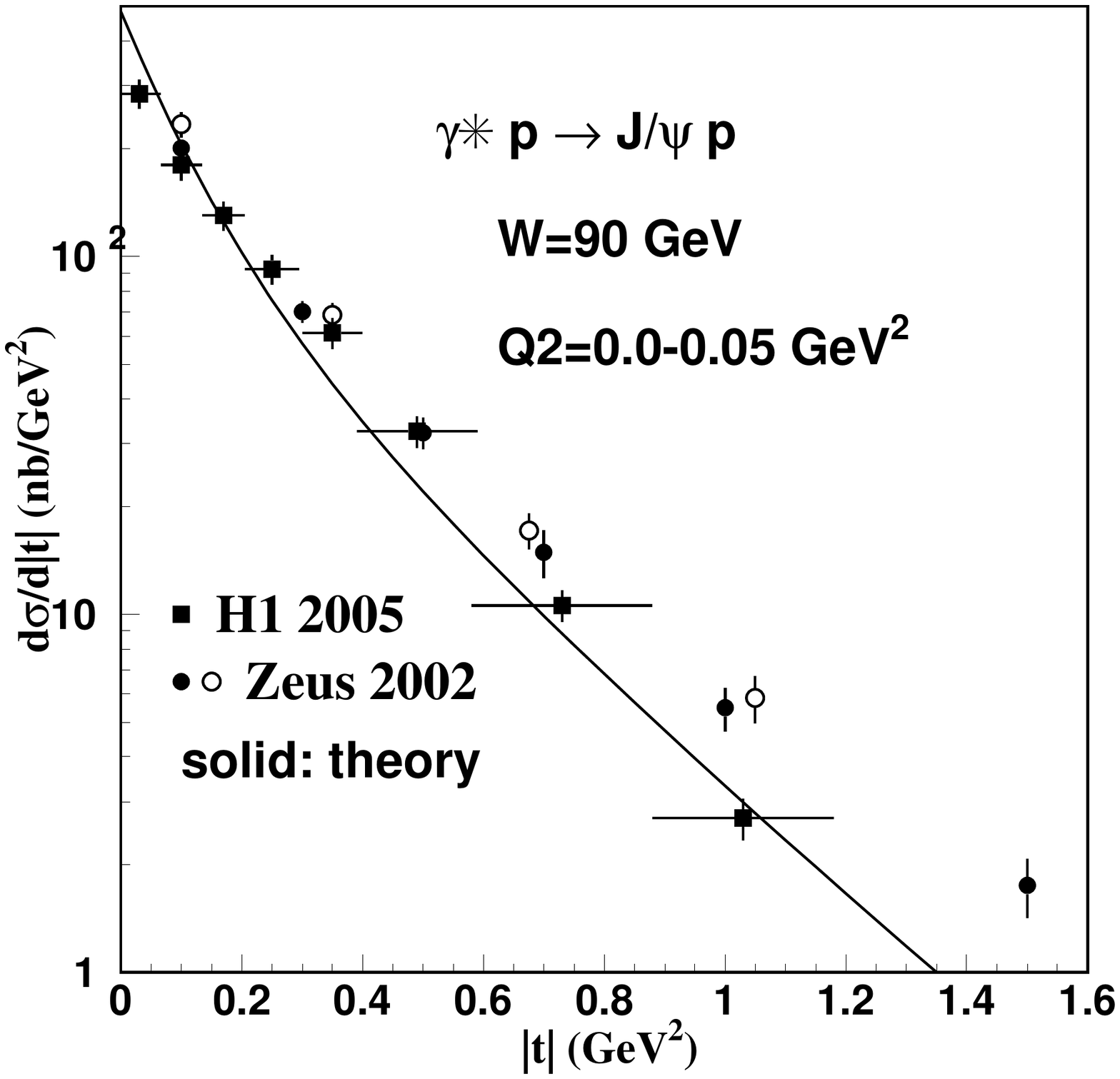}
 \includegraphics[height=7cm,width=7cm]{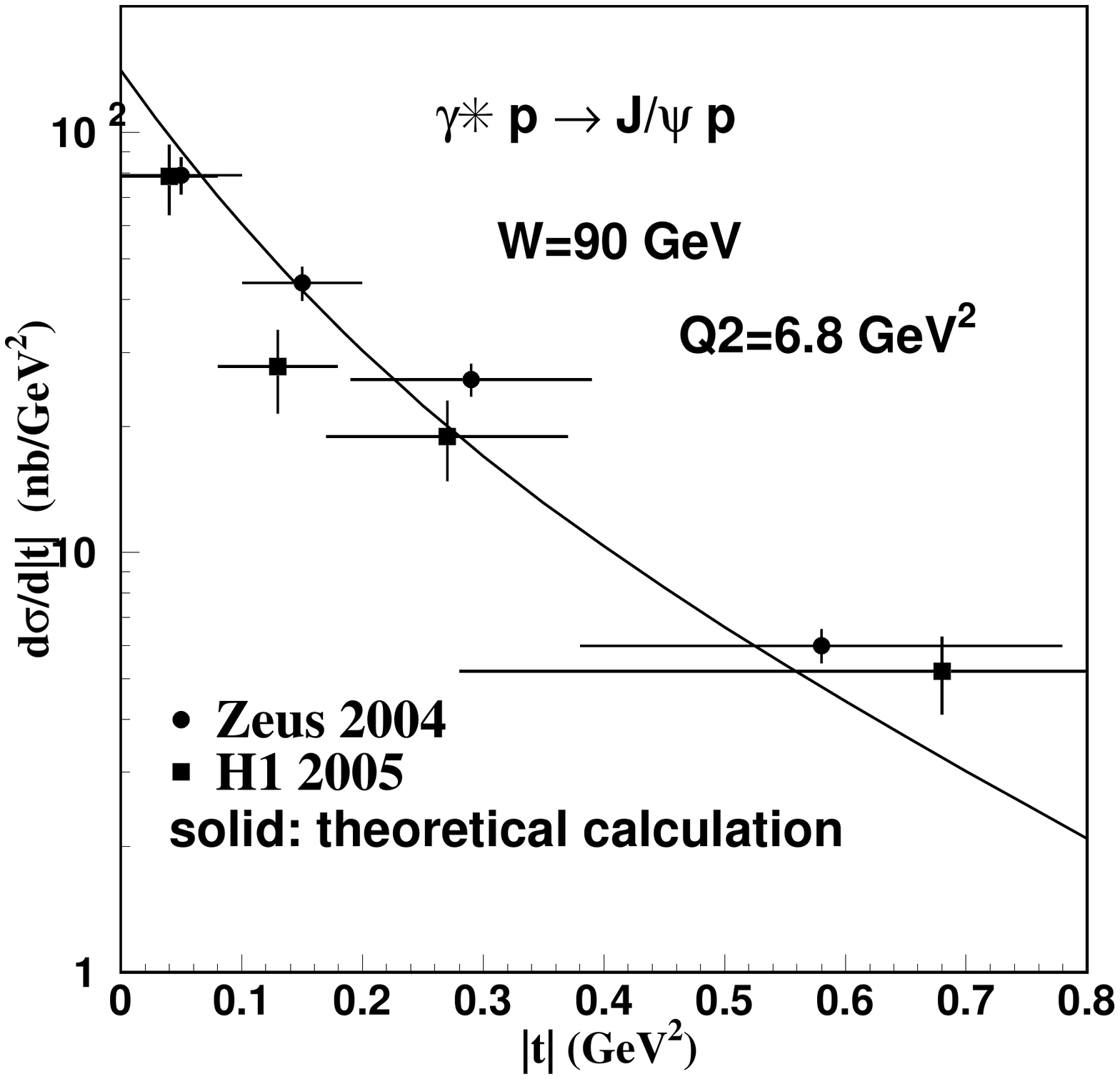}
 \caption{\label{tpsi} The $t$ dependence of the
differential cross sections of J/$\psi$ elastic photo
\cite{Zeus2002,H12005} and electroproduction \cite{Zeus2004,H12005}.
The Zeus(2002) photoproduction results
have  separate information from events with $\psi$ decaying into
$\mu^+\mu^-$ (full circles) and $e^+e^-$ (empty circles).
 The solid lines represent our  theoretical results.}
\end{figure}

 In contrast to purely perturbative approaches, our theoretical treatment
allows to calculate the dependence of the cross section on the
momentum transfer $t$. The data on the $t$-distribution in
$J/\psi$ photo \cite{Zeus2002,H12005} and electroproduction
\cite{Zeus2004,H12005} at W=90 GeV are  shown in Fig. \ref{tpsi}
together with our theoretical results (solid lines). Also shown
are electroproduction data and calculations at $Q^2=6.8-7 $
GeV$^2$ where measurements from both Zeus and H1 are available.

Our calculation predicts a curvature in the log plot for the
$t$ distribution, which we may describe with the form
\begin{equation} \label{ff_eq}
\frac{d\sigma}{d|t|}=\Bigg[\frac{d\sigma}{d|t|}\Bigg]_{t=0}
\times F(|t|)=
\Bigg[\frac{d\sigma}{d|t|}\Bigg]_{t=0}
\times  \frac{e^{-b|t|}}{(1+a|t|)^2}~ .
\end{equation}
For  $\gamma^\star p \rightarrow p \psi$ at W=90 GeV our
calculations at $Q^2=0$ give   $a=4.06 ~ {\rm GeV}^{-2}$ and
$b=1.75 ~ {\rm GeV}^{-2}$. Present data give no clear cut evidence
for this curvature.  The distribution becomes flatter as $Q^2$
increases, but in the investigated range does  practically not
change with the energy (no shrinking).

In order to determine the integrated cross section the
experimental data have to be extrapolated to the point of minimum
momentum transfer. This brings a theoretical bias to the
experimental points. It is therefore meaningful to compare our
theoretical results with observed data at small but finite
momentum transfer. This is done in   Fig. {\ref{psiforward};  the
agreement is excellent.

\begin{figure}[h]
 \vskip 2mm
 \includegraphics[height=7cm,width=7cm]{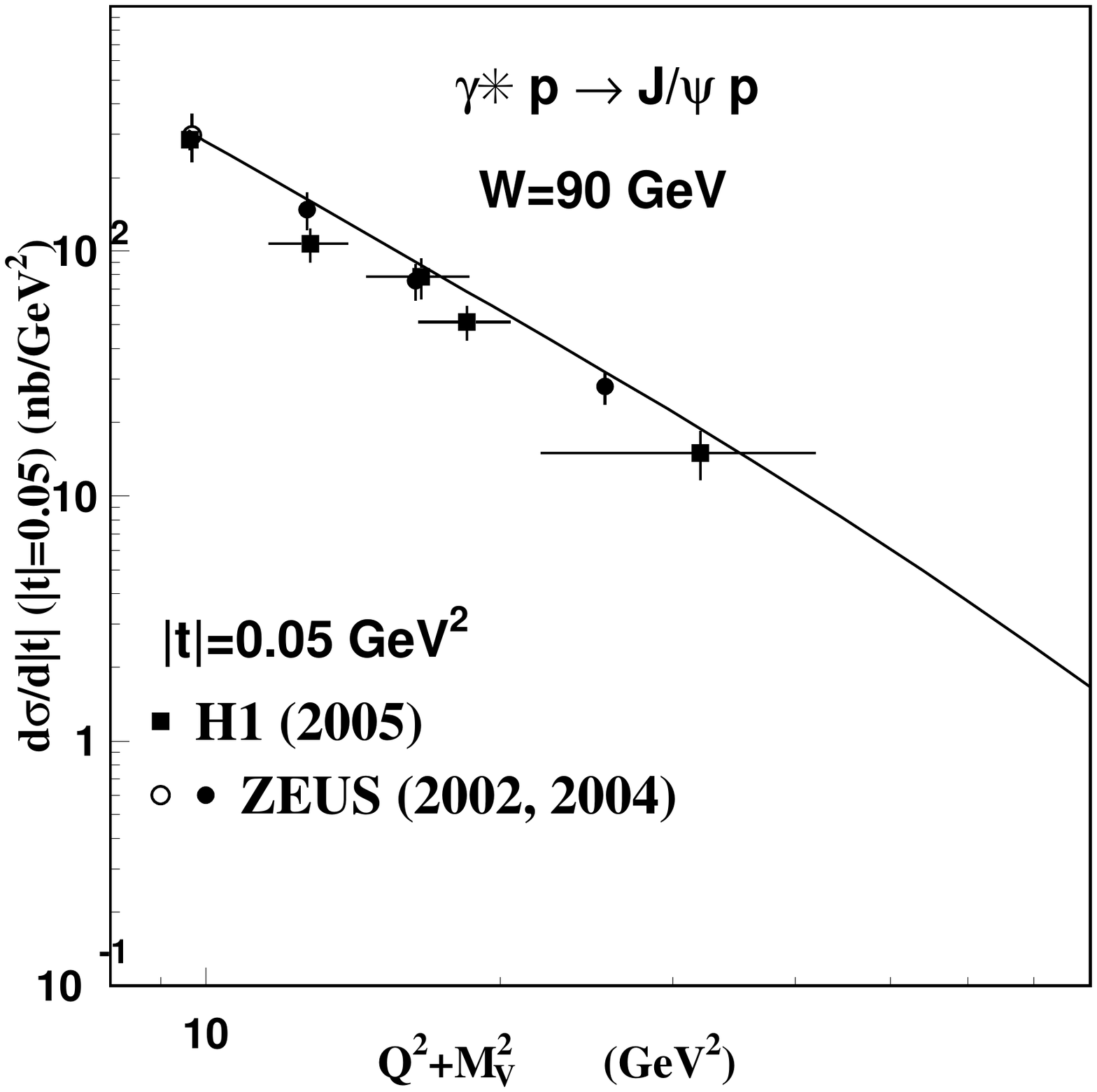}
 \includegraphics[height=7cm,width=7cm]{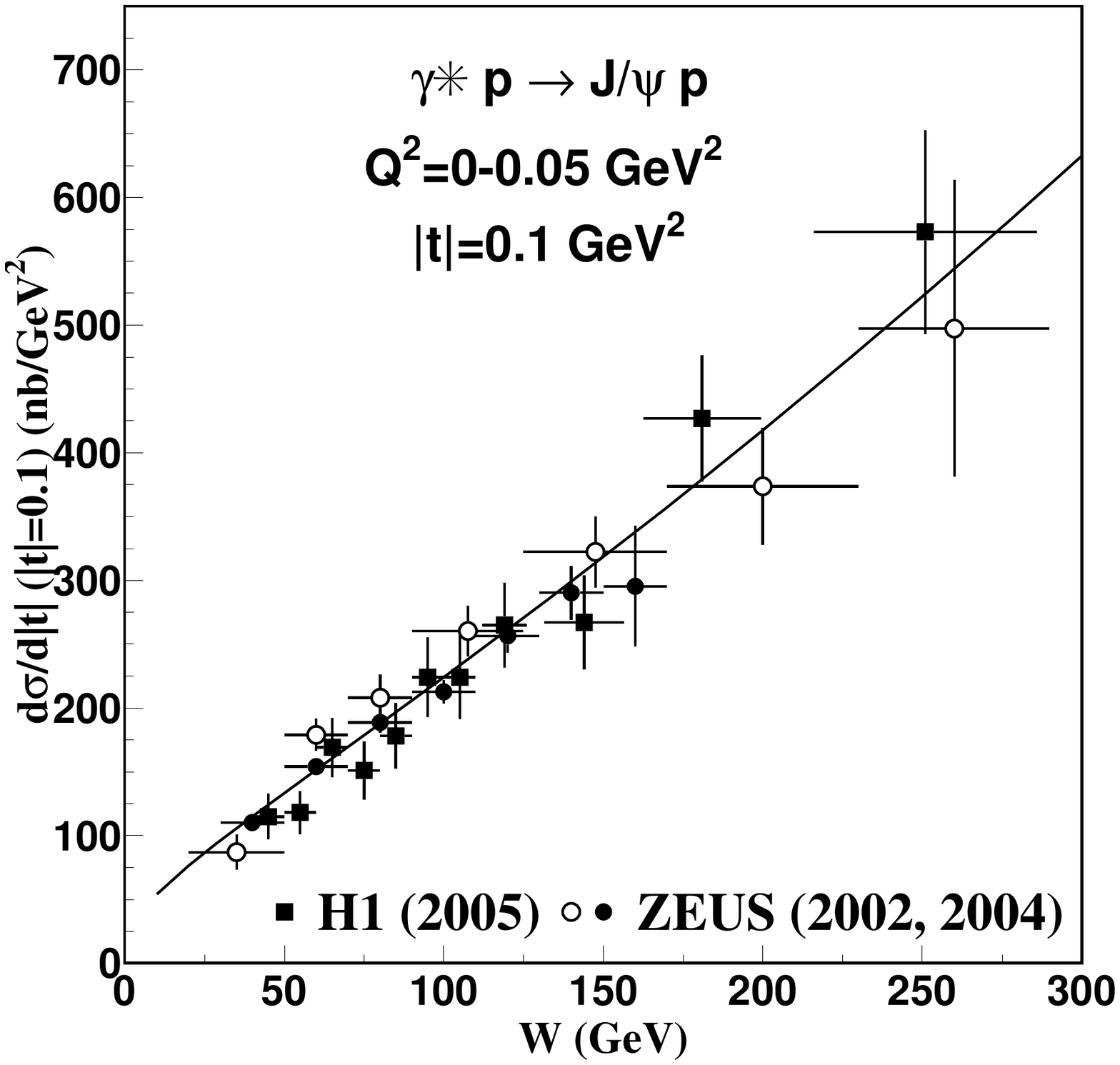}
 \caption{\label{psiforward} $Q^2$ and W dependences of the
 (nearly) forward differential cross sections of J/$\psi$ elastic photo
\cite{Zeus2002,H12005} and electroproduction \cite{Zeus2004,H12005}.
The solid lines represent our  theoretical results.}
\end{figure}

\subsection{Photo and electroproduction of $\Upsilon$ meson}

For the $\Upsilon$ there are only two data points for photoproduction,
at $<W>$=120 GeV \cite{Zeus98b}and $<W>$=143 GeV \cite{H1_PLB2000}, shown together with our theoretical   energy dependence
in Fig. \ref{ups}. In the second plot we show our calculation of the
$Q^2$ dependence at the fixed energy W=130 GeV, together with the data for $Q^2=0$ at the energies 120 and 143 GeV.   
Both plots show a fairly good agreement with the data.
It should be noted that, in spite of the high mass scale, the finite
extension of the meson has a considerable effect, as can be seen in
Fig. \ref{extdip},left.

 \begin{figure}[h]
 \vskip 2mme
 \includegraphics[height=7cm,width=7cm]{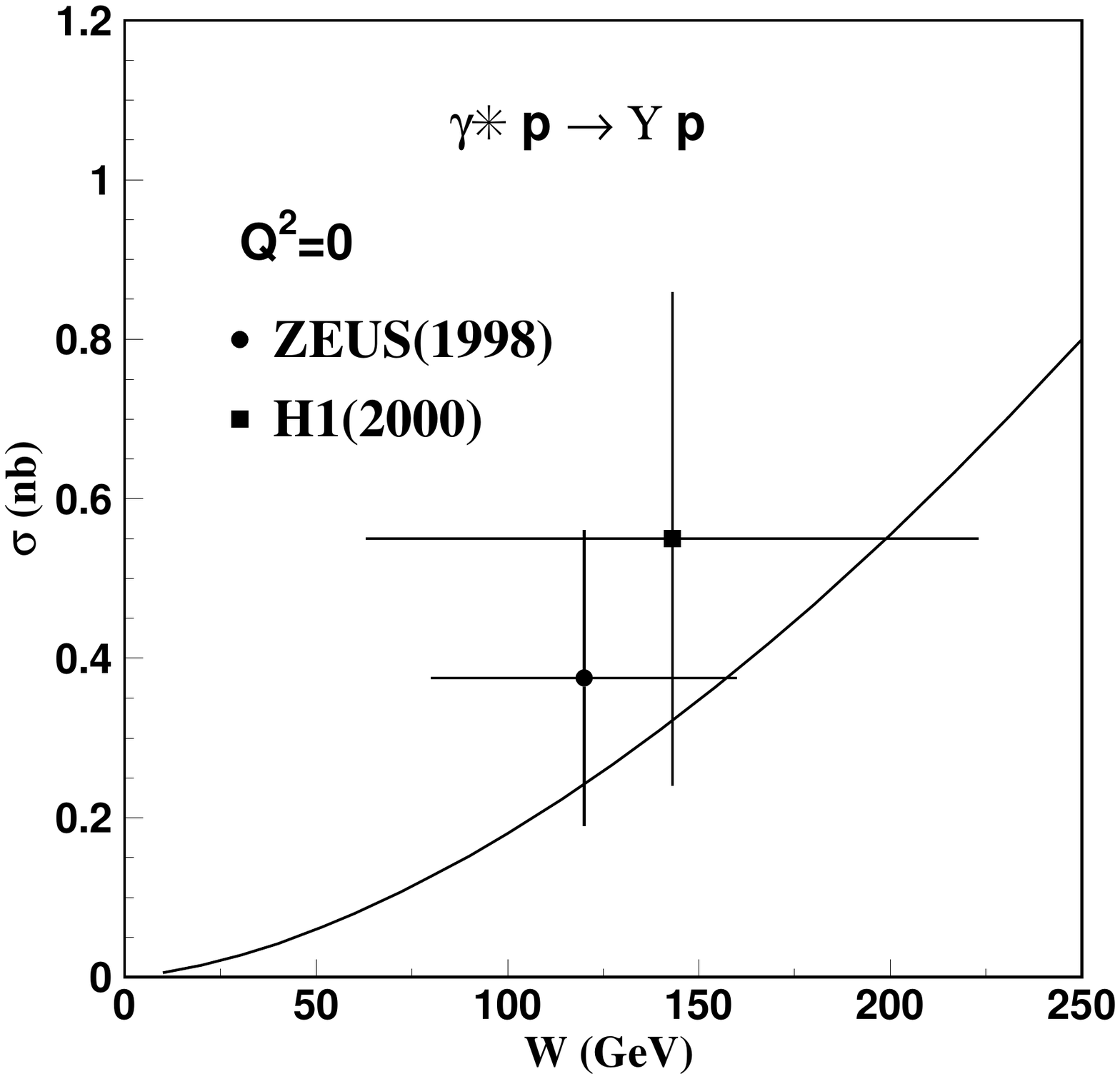}
 \includegraphics[height=7cm,width=7cm]{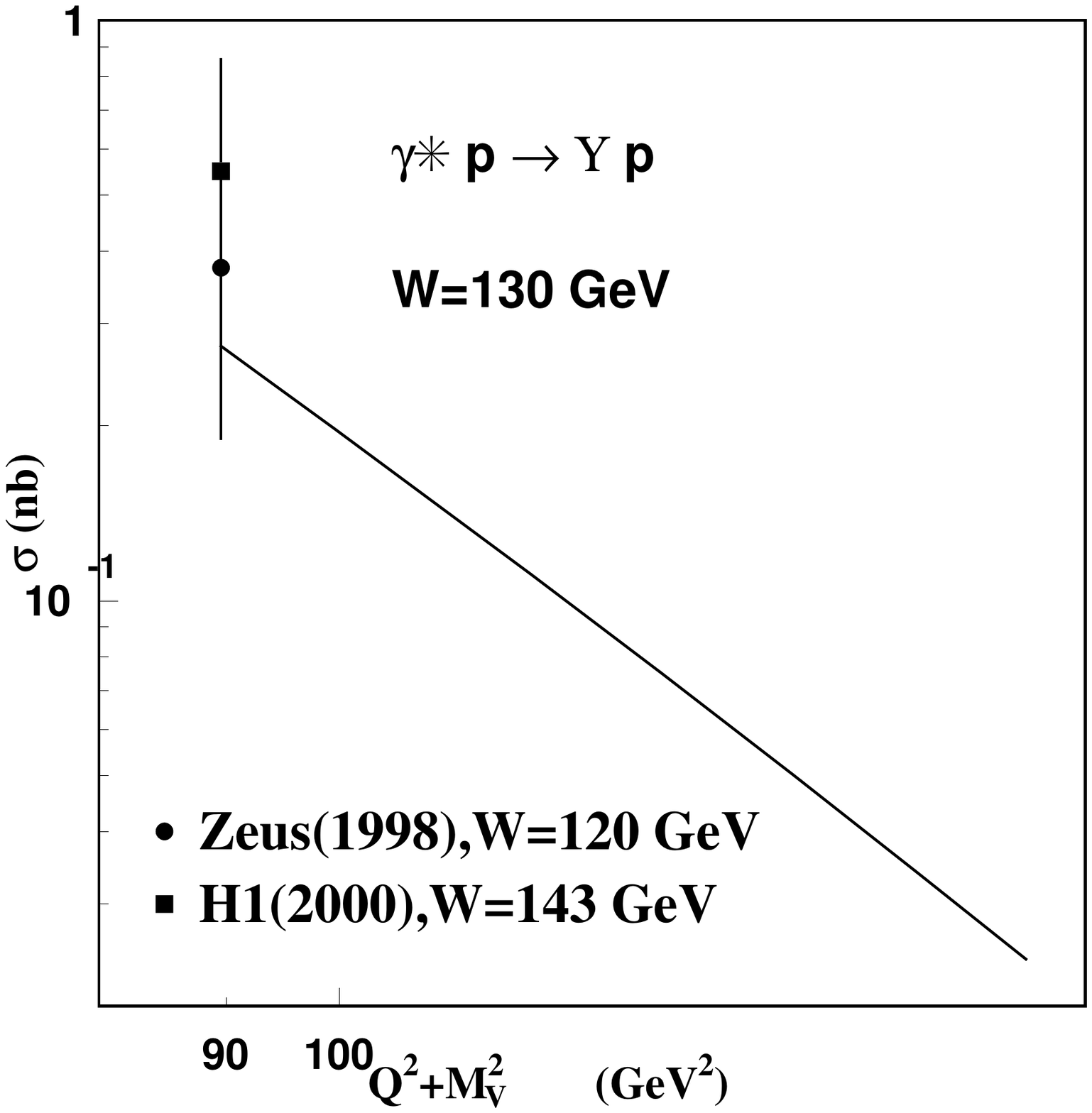}
 \caption{\label{ups} Integrated elastic cross
section of $\Upsilon$ photoproduction  as function of the energy
and electroproduction at the energy W=130 GeV as a function of $Q^2$.
The data at the energies $<W>$ = 120 and 143 GeV are respectively
from  Zeus \cite{Zeus98b} and H1 \cite{H1_PLB2000} collaborations.
The solid lines represent our theoretical calculations using the BL
wave function.}
\end{figure}

\subsection{Photo and electroproduction of $\rho$ meson}

The classical fixed target experiments \cite{egloff,chio,emc,e665,NMC94}
 provide important reference
for the  magnitudes of cross sections in $\rho$ photo and
electroproduction  at center of mass energies about  20 GeV. The
data are shown in Fig. \ref{rhosig20}, together with the results
of our calculation.  The agreement is satisfactory given the quite
important discrepancies within the data in the interval $Q^2=4-7
\GeV ^2$.

 \begin{figure}[h]
 \vskip 2mm
 \includegraphics[height=8cm,width=8cm]{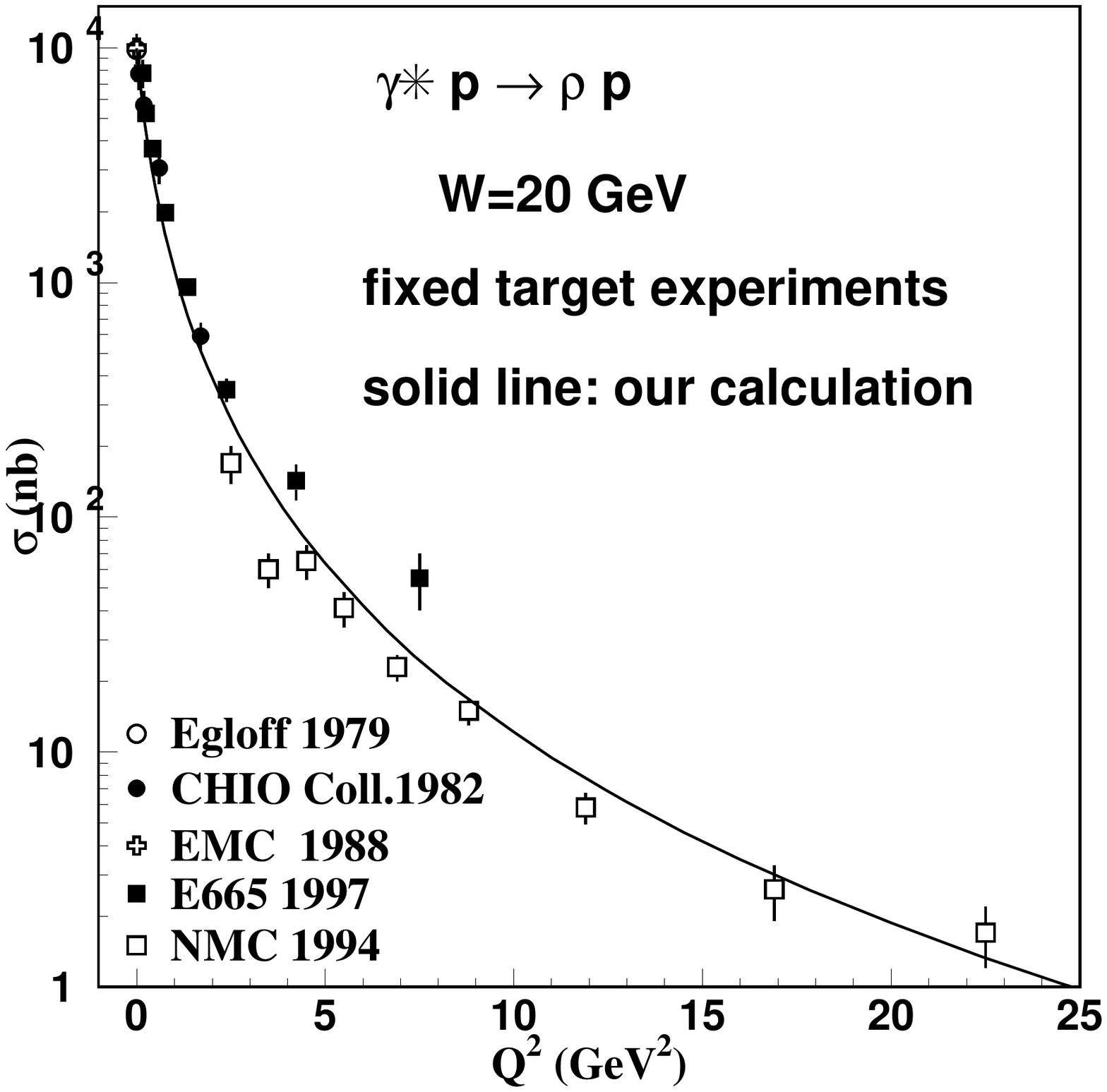}
 \caption{\label{rhosig20} Integrated elastic cross
section for  $\rho$  electroproduction  at the energy W=20 GeV as
a function of the photon virtuality $Q^2$. The data are from fixed
target experiments \cite{egloff,chio,emc,e665,NMC94}, spanning
almost two decades. The solid line represents the results of our
calculations.}
 \end{figure}

Fig. \ref{rhoq2} shows the data for the integrated elastic cross
section $\sigma$ of $\rho$ electroproduction at the fixed energy
W=90 GeV as a function of the photon virtuality $Q^2$. The data
are from Zeus \cite{Zeus99,Zeus2001} and H1 \cite{H12000,H12003}
collaborations. The solid line represents the theoretical
calculations using the BL form of the vector meson wave function.
Our theoretical calculations give an good overall description of
the data but do not reproduce the details of   the $Q^2$
dependence of $\rho$ electroproduction as  well as they did in
J/$\psi$ production. The present data indicate that in the $\rho$
meson case a fitting of a simple form like Eq. (\ref{psiq2fit}) is
not satisfactory for the whole $Q^2$ range. Two of such forms may
be applied, with a transition in the values of both parameters
(normalization and power) occurring somewhere in the  $Q^2$ range
from 5 to 7. More experimental measurements are needed to clarify
the structure of the data in this region, which may contain
important information about the dynamics of the process.

 \begin{figure}[h]
 \vskip 2mm
 \includegraphics[height=8cm,width=8cm]{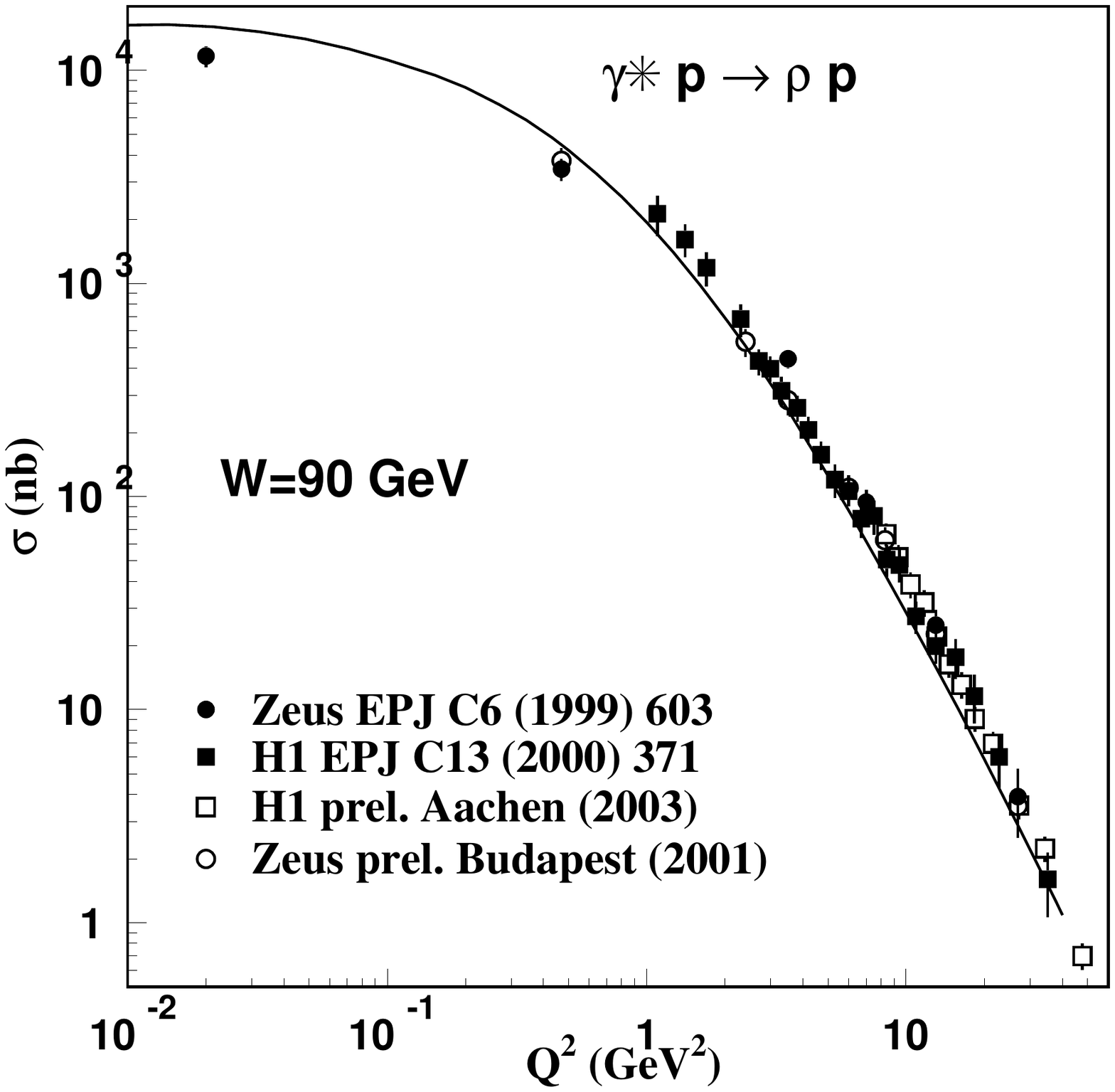}
 \caption{\label{rhoq2} Integrated elastic cross section of
$\rho$  electroproduction  at the energy W=90 GeV as a function
of $Q^2$. The data are from Zeus \cite{Zeus99,Zeus2001}
and H1 \cite{H12000,H12003} collaborations. The solid line
shows our theoretical results using the BL wave function.}
\end{figure}

Fig. \ref{rhoratio} shows the ratio $ R=\sigma^L/\sigma^T $ of
cross sections for  longitudinal and transverse polarisations, for
$\rho$ elastic electroproduction at W=20 and 90 GeV.  The data at
90 GeV are from NMC \cite{NMC94}, Zeus \cite{Zeus99,Zeus2001} and
H1 \cite{H196,H12000,H12003}. The W=20 GeV data are from the E-665
experiment \cite{e665}. To indicate the differences, the
theoretical results for both BL (solid line) and BSW (dashed line)
wave functions are displayed. For large $Q^2$,  $R$ becomes very
sensitive to the values of the small transverse cross section. The
measurement of this quantity $R$ therefore provides important tests
on the structure of the meson wave function.

 \begin{figure}[h]
 \vskip 2mm
 \includegraphics[height=7cm,width=7cm]{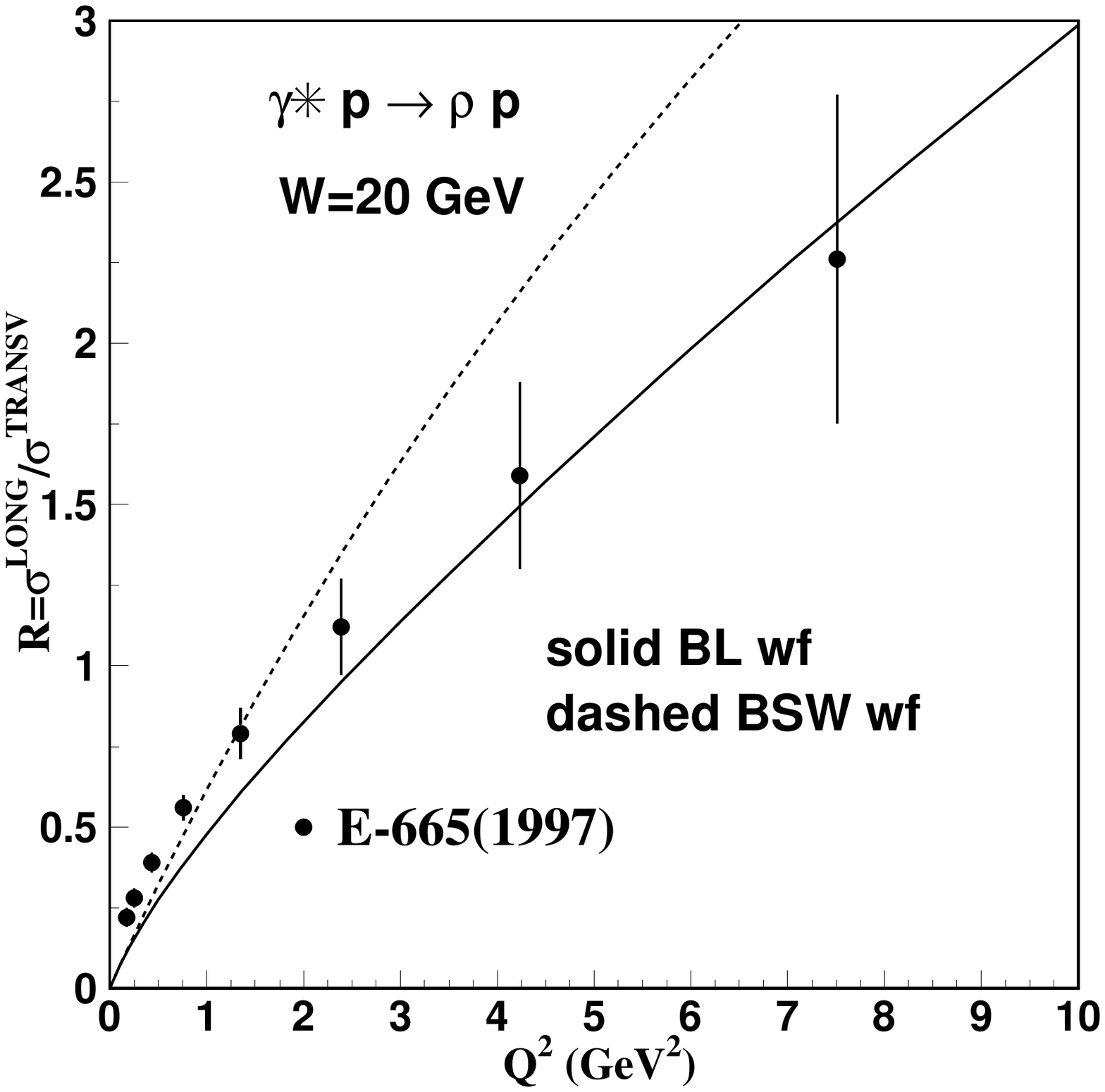}
\includegraphics[height=7cm,width=7cm]{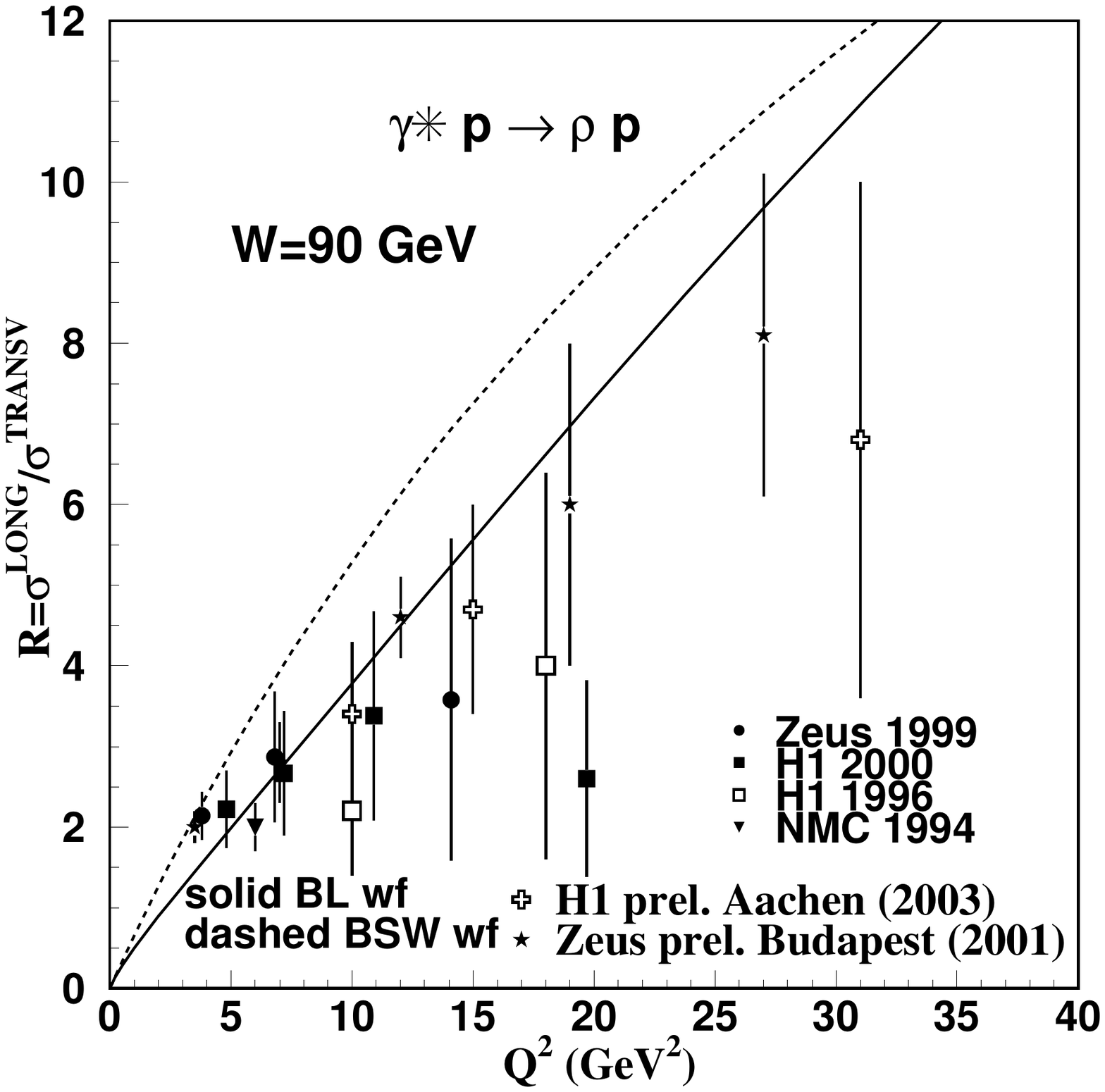}
 \caption{\label{rhoratio} Ratio R between longitudinal and
transverse cross sections for $\rho$ electroproduction as
function of $Q^2$ for fixed energies W=20 GeV , with data from
E-665 \cite{e665}, and W=90 GeV with data from  NMC \cite{NMC94},
Zeus \cite{Zeus99,Zeus2001} and H1 \cite{H196,H12000,H12003}.
The solid and dashed lines show the theoretical calculations
with the BL and BSW wave functions respectively.}
\end{figure}

Fig. \ref{rhoW} shows the energy dependence of $\rho$
electroproduction for several values of $Q^2$ together with our
theoretical results. On the right-hand-side plot we show  the
effective power $\delta$  for the energy dependence (see eq.
(\ref{wpower})) for the $W$-region of approximately 30 to 130 GeV.
The experimental points  are from Zeus
\cite{Zeus99,Zeus2001,Zeus98} and H1 \cite{H12000,H12003}. Our
theoretical description, solid line,  is very satisfactory within
the experimental errors.

 \begin{figure}[h]
 \vskip 2mm
 \includegraphics[height=7cm,width=7cm]{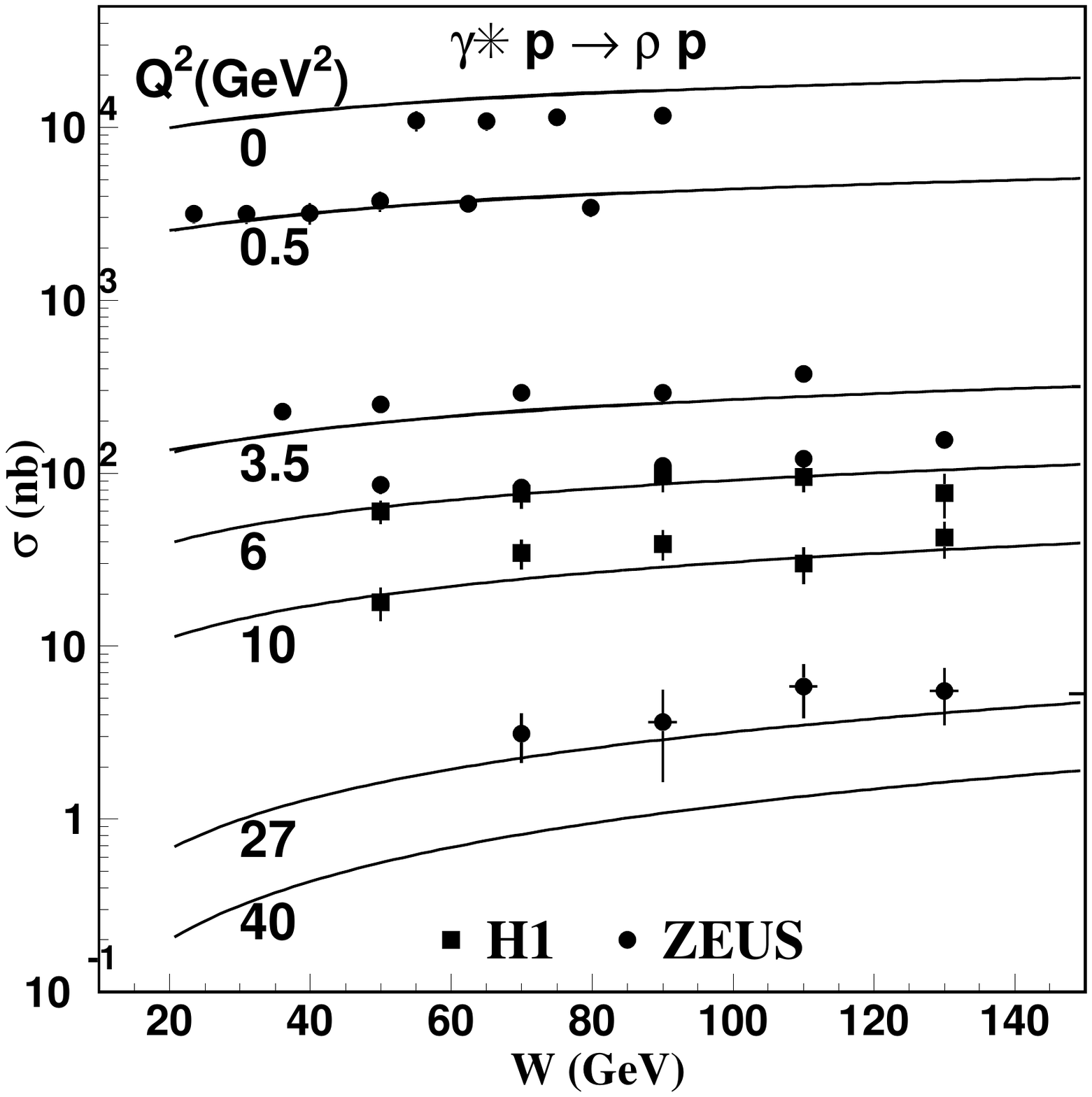}
 \includegraphics[height=7cm,width=7cm]{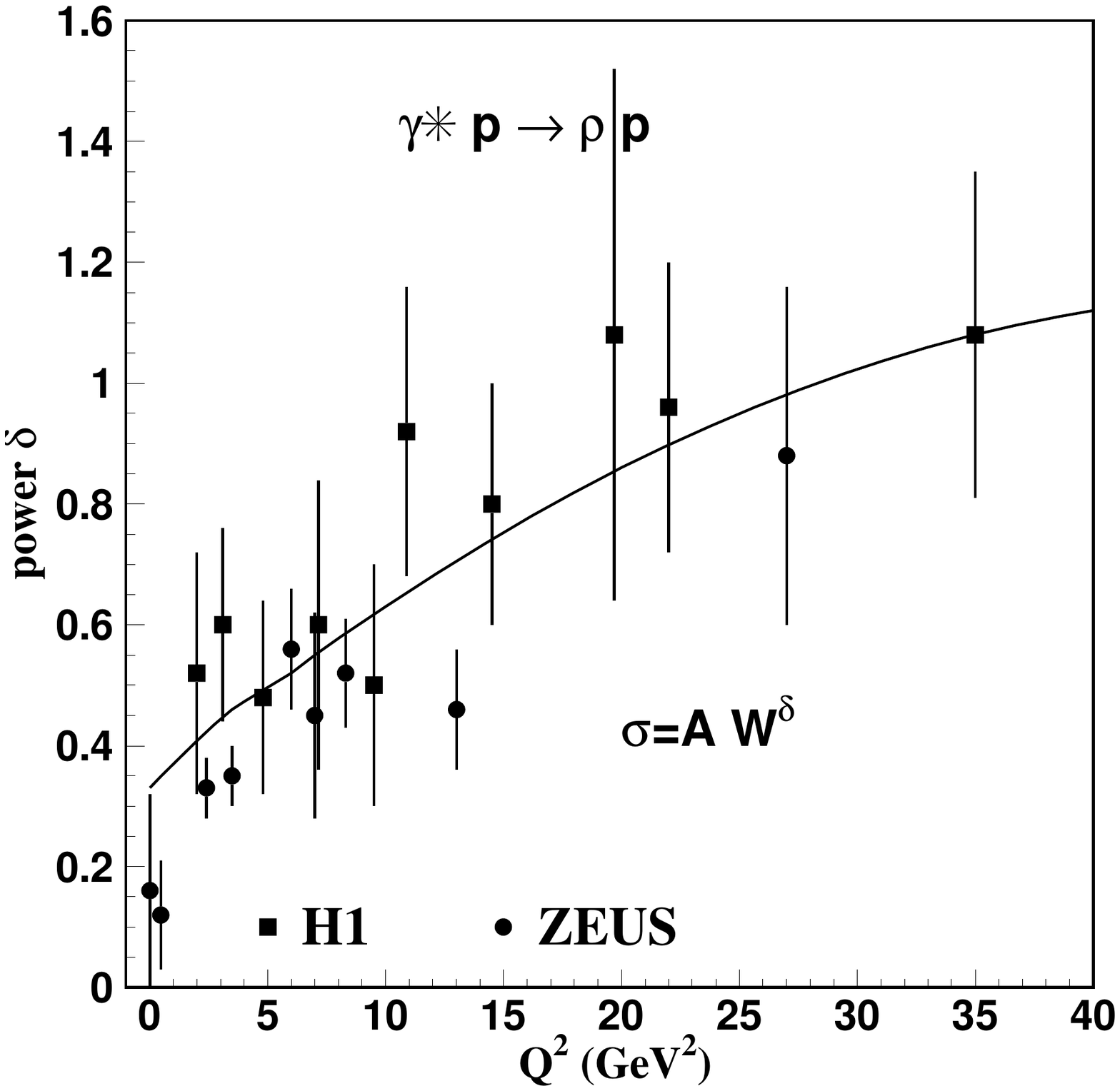}
 \caption{\label{rhoW} Energy dependence of $\rho$ electroproduction
cross section in the region from approximately 30 to 130 GeV. The
solid lines show our theoretical calculations for several values
of $Q^2$ and the data are from  Zeus \cite{Zeus99,Zeus2001,Zeus98}
and H1 \cite{H12000,H12003} collaborations. The plot in the right
hand side shows data and theoretical values for the parameter
$\delta$ of the energy dependence $W^\delta$. }
\end{figure}

 There are no published measurements of the differential cross
section $d\sigma/d|t|$ in $\rho$ production for nonzero values of
$Q^2$. The Zeus photoproduction data at W=75 and 94 GeV
\cite{Zeus98,EPJC14}  are shown in Figs. \ref{rhot}. In the low
$t$ range there are new preliminary data from H1
\cite{H1_prelim2006} at several energies. The numbers for the
differential cross sections at 70 GeV for low t extracted from
their plots are included in the figure, and they seem to confirm
the previous Zeus measurements at W=75 GeV \cite{Zeus98}. Our
theoretical calculations are shown in the figures, the description
is satisfactory.

 \begin{figure}[h]
 \vskip 2mm
 \includegraphics[height=7cm,width=7cm]{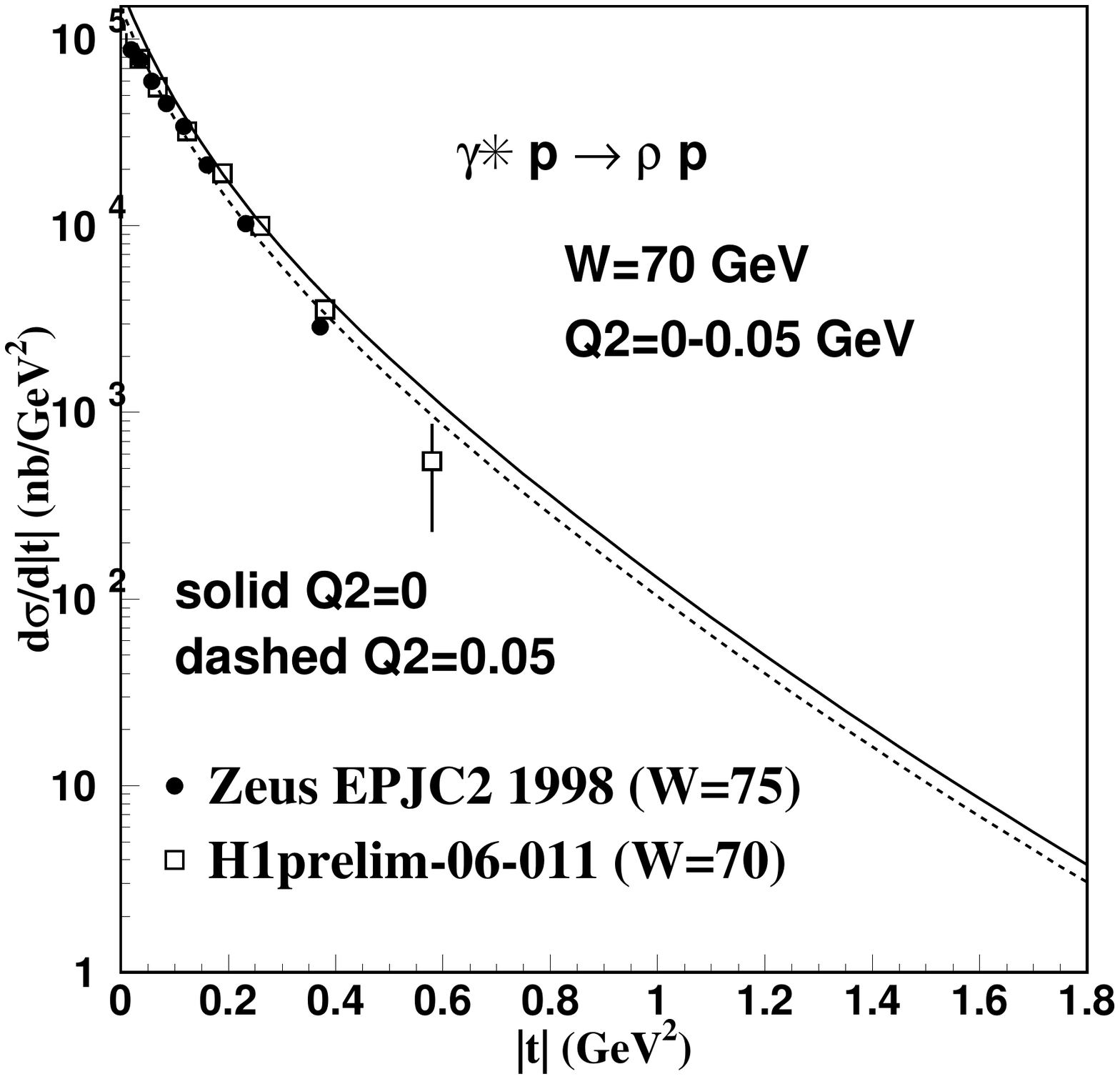}
 \includegraphics[height=7cm,width=7cm]{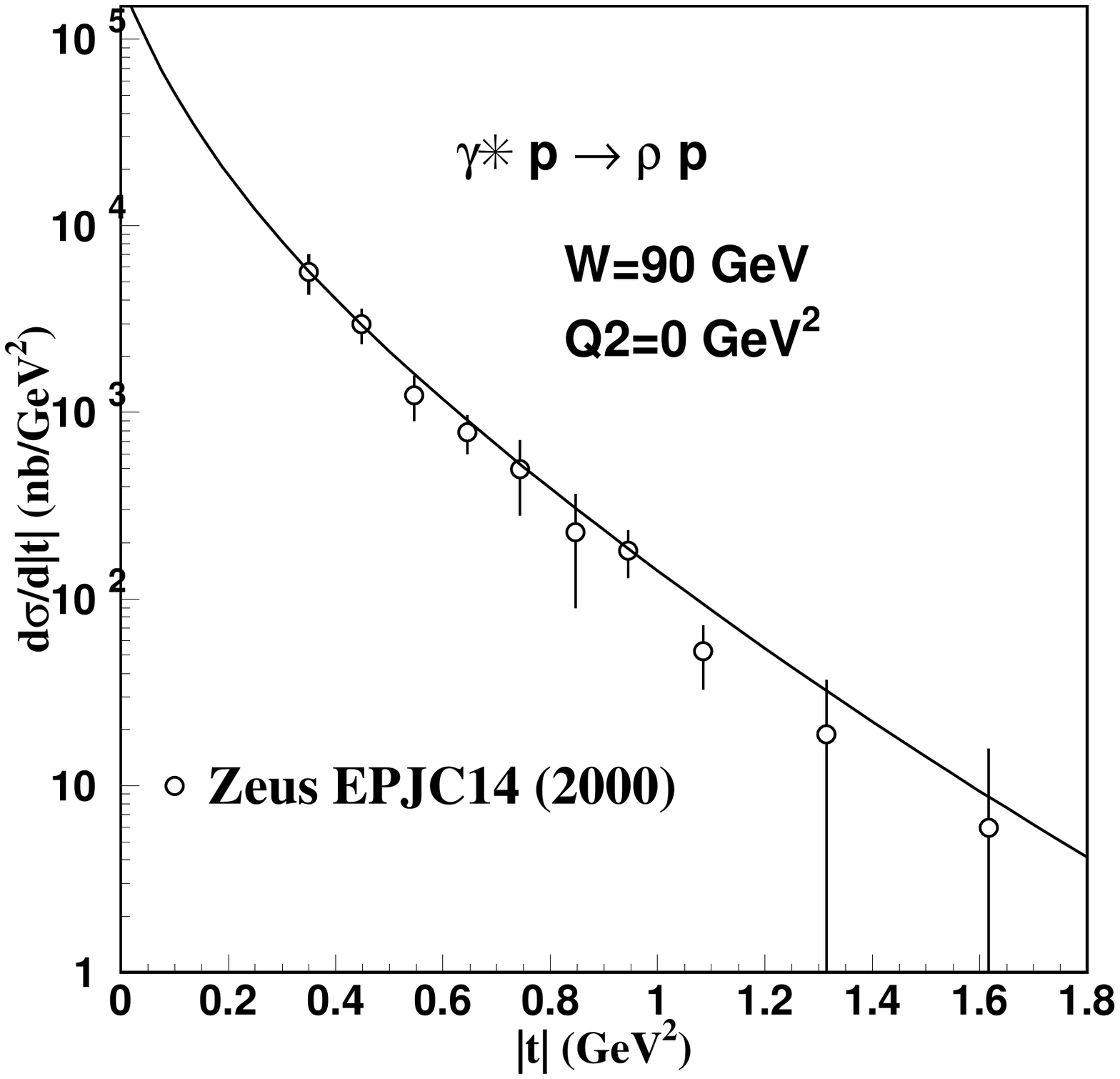}
 \caption{\label{rhot} Experimental data and our theoretical 
prediction (solid line) for the $t$ dependence of $\rho$ 
photoproduction  cross sections.
The published data are from  the Zeus collaboration, at the
energies 75 GeV  (low t) \cite{Zeus98} and 94 GeV (large t)
\cite{EPJC14}. The 94 GeV data  are rescaled with  a factor
$(90/94)^{0.16}=0.993$ in the figure. We also  include in the
low-t figure  preliminary information from H1
\cite{H1_prelim2006}, extracted
 from their plots, for  W=70 GeV. }
 \end{figure} 
 
\subsection{Photo and electroproduction of $\omega$ meson}

As particles of about the same size and mass, $\omega$ and $\rho$
have similar behaviour in the soft processes that we study here.

The $Q^2$ dependence of $\omega$ photo and electroproduction
\cite{omega96,omega2000} is shown in Fig. \ref{omegaq2}. The
agreement between our results and the data  is not perfect, but
satisfactory, in view that there is no free parameter involved in
the calculations.

 \begin{figure}[h]
 \includegraphics[height=8cm,width=8cm]{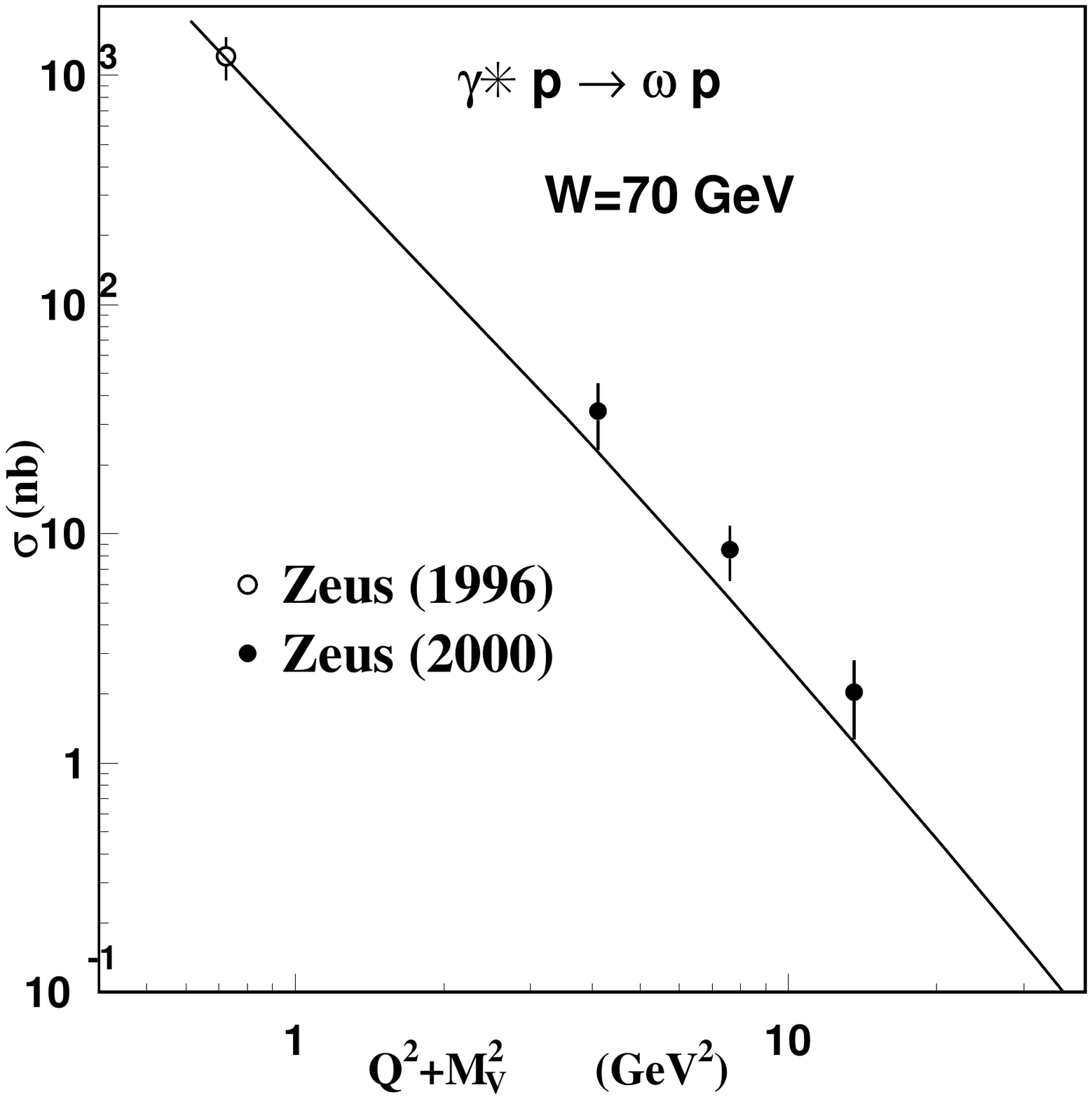}
 \caption{\label{omegaq2} $Q^2$ dependence of the  cross section
of $\omega$ elastic production \cite{omega96,omega2000}. The solid
line represents our calculations. }
\end{figure}

The existing data \cite{Busenitz89,Derrick96} on the $t$
dependence of the differential cross section in $\omega$
production are shown in Fig. \ref{t_omega}. In the figure are put
together the data points of the energies W=15 GeV (with $Q^2=0$)
and  W=80 GeV (with $Q^2$=0.1 GeV$^2$), and the corresponding
theoretical curves. The coincidence of shapes exhibits the universality
of the form factors of $t$ dependence in our model. Our prediction
of a curvature in the plot of  $d\sigma/d|t|$ seems to be confirmed
by the data.
 \begin{figure}[h]
 \vskip 2mm
 \includegraphics[height=8cm,width=8cm]{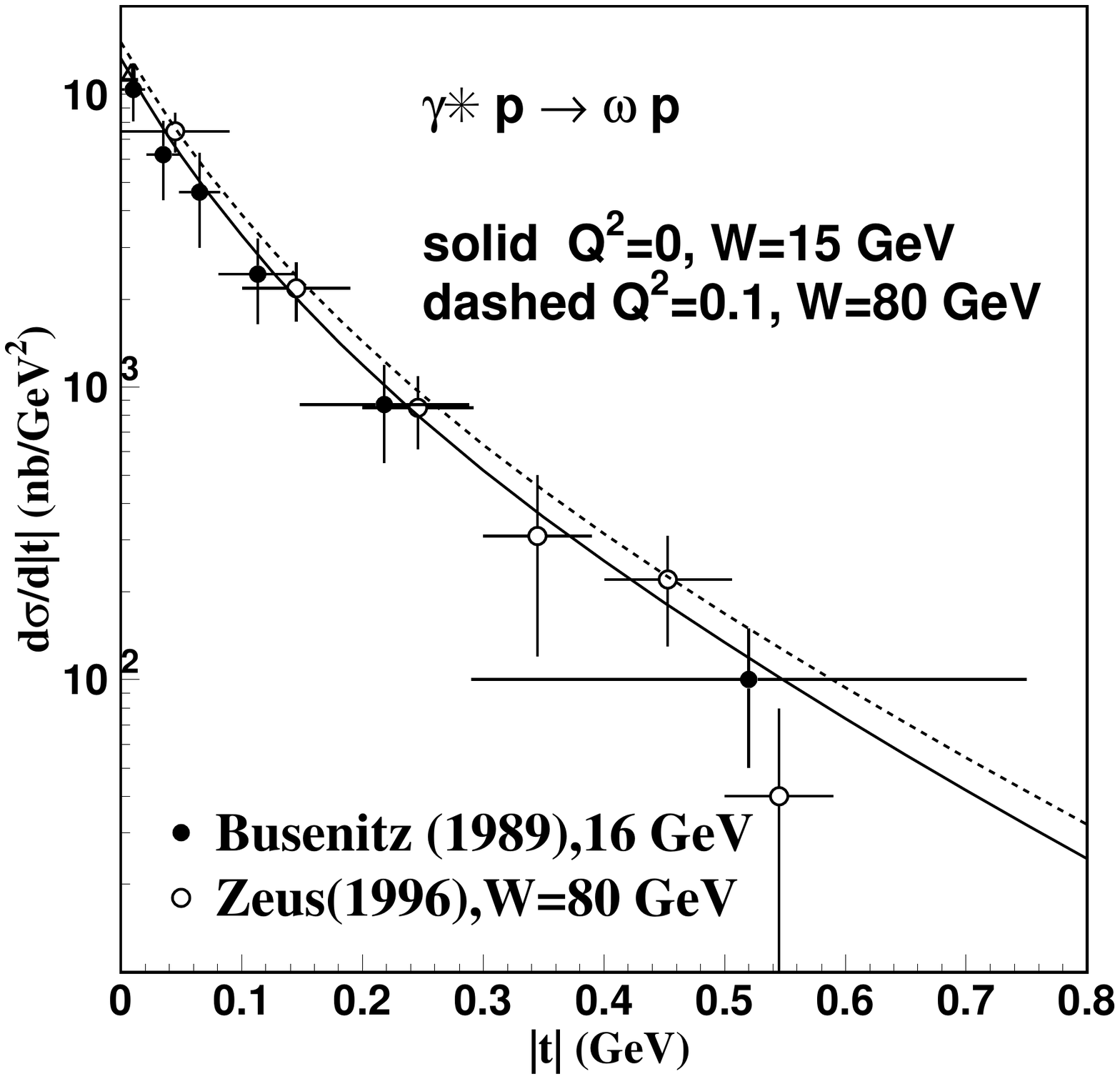}
  \caption{\label{t_omega} t dependence of $\omega$ photoproduction
 cross sections    \cite{Busenitz89,Derrick96}.
 The plot puts  together data points at the energies W=15 GeV
(with $Q^2=0$) and  W=80 GeV (with $Q^2$=0.1 GeV$^2$), and the
corresponding theoretical curves. The curvature and the
similarities of shapes of the $t$ distributions
 for two different energies and $Q^2$ values   are characteristic
 features of our framework. }
 \end{figure}

\subsection{Photo and electroproduction of $\phi$ meson}

The $\phi$ meson has a strategic place between the $J/\psi$ and the
$\rho$ meson and may help to understand the differences in
behaviour of heavy and light vector mesons and also  to clarify
the interplay of perturbative and nonperturbative aspects of QCD.

In Fig. \ref{phiq2} we show the $Q^2$ dependence of the integrated
elastic cross section in $\phi$ photo and electroproduction at
W=75 GeV \cite{phi96,phi2000,phi2005}, together with our
theoretical calculations. As in the case of $\rho$
electroproduction, there is indication that the data cannot be
well represented by a single expression of the form of
Eq. (\ref{psiq2fit}). The data are poorer here than in the
$\rho$ case, and it is important to investigate the possibility
of a transition region in an intermediate $Q^2$ range below 10
GeV$^2$ in which the normalization and the power change values
rather rapidly.

\begin{figure}[h]
 \vskip 2mm
 \includegraphics[height=8cm,width=8cm]{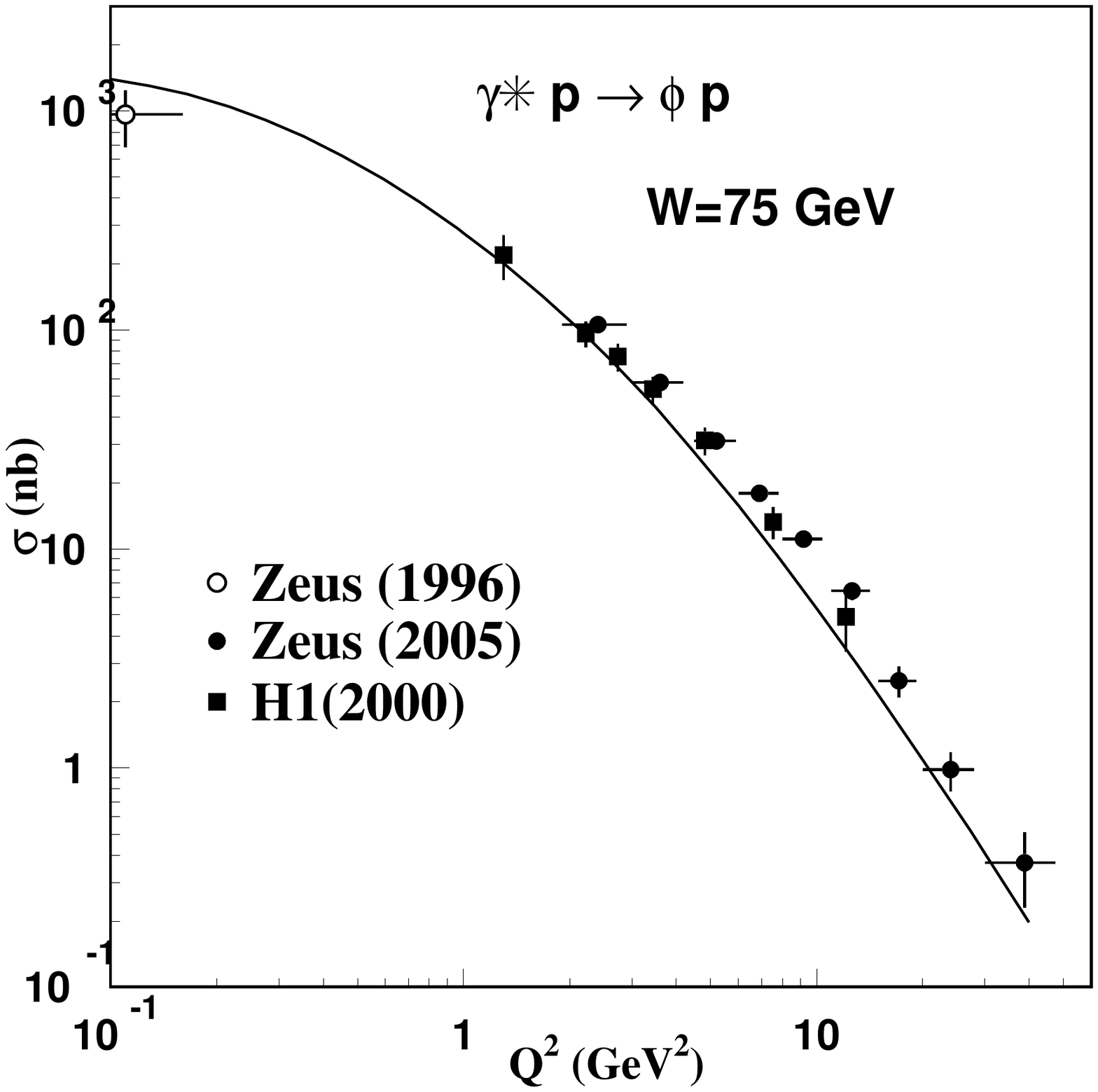}
 \caption{\label{phiq2} $Q^2$  dependence of $\phi$ production
 cross sections at W=75 GeV and our theoretical  description.
 The  data are from Zeus \cite{phi96,phi2005} and
 H1 \cite{phi2000}.}
 \end{figure}

  The data for the ratio $R=\sigma^L/\sigma^T$ for W=75 GeV as
function of  $Q^2$ \cite{phi2000,phi2005}  are shown in Fig.
\ref{phiratio}, together with our results with BL and BSW wave
functions. The calculations exhibit, as in the $\rho$ meson case,
the sensitivity of the ratio $R$ to details of the wave functions.
Offering a reference for the two kinds of calculation, the plot
also shows (dotted line) a fit of the data, made by
experimentalists.

\begin{figure}[h]
 \vskip 2mm
 \includegraphics[height=8cm,width=8cm]{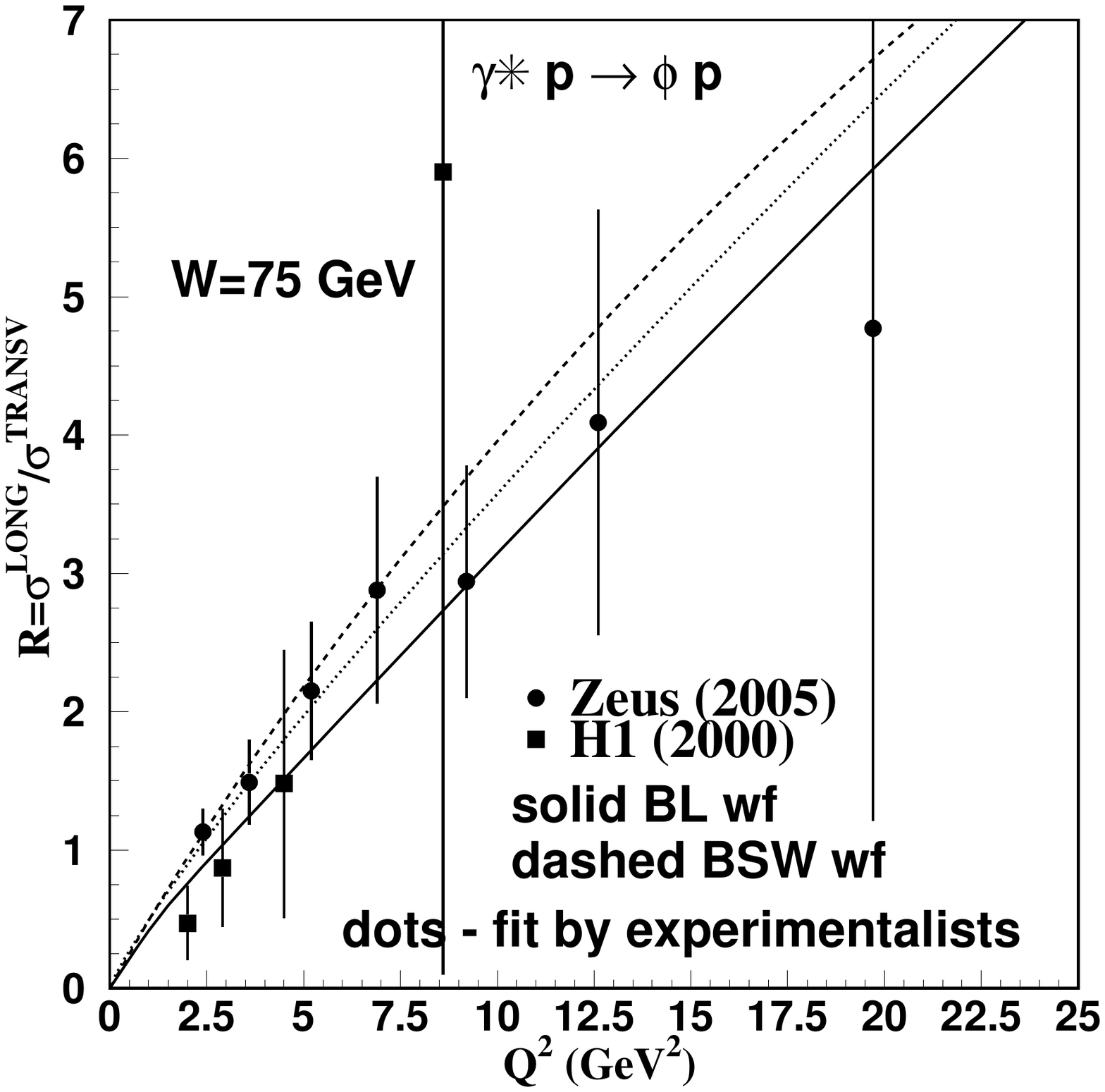}
 \caption{\label{phiratio}
 $Q^2$ dependence of the ratio
$R=\sigma^L/\sigma^T$ in  $\phi$ electroproduction
\cite{phi2000,phi2005} at fixed energy W=75 GeV.  The solid and
dashed lines represent our calculations with BL (solid) and BSW
(dashed) wave functions; the dotted line is a  fit  of the form
 $R=0.51~ (Q^2/M^2)^{0.86}$.}
 \end{figure}

The effective power  $\delta(Q^2)$  describing the energy
dependence as $W^\delta$  in $\phi$ electroproduction
\cite{phi2005} is shown in Fig. \ref{phidelta}, and compared with
our results using the two-pomeron model. The theoretical $Q^2$
dependence is similar to that of $J/\psi$ and $\rho$ production,
whereas the experimental points, with large errors, indicate a
flatter behaviour. Obviously more data are necessary.

 \begin{figure}[h]
 \vskip 2mm
  \includegraphics[height=8cm,width=8cm]{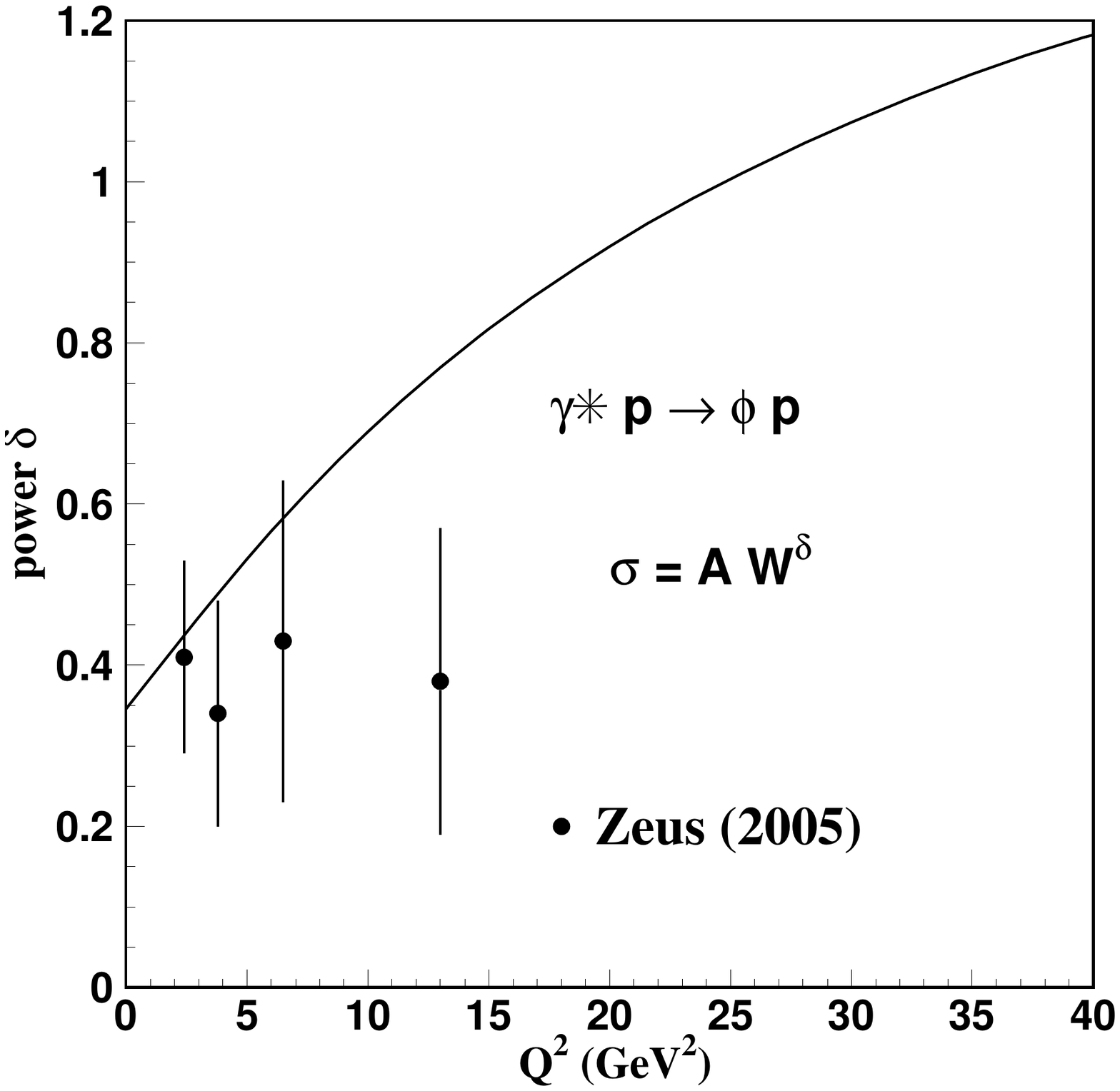}
 \caption{\label{phidelta}
 The effective power $\delta$ parameter governing the energy
dependence of $\phi$ electroproduction. The data are from Zeus
\cite{phi2005}, the solid line is our theoretical result.}
\end{figure}

Our calculations give very good descriptions for the $Q^2$
dependence  of the differential elastic cross section in
forward directions in $\phi$ electroproduction, as it does
in the $J/\psi$ case. Fig. {\ref{phiforward} shows the
comparison with the experimental data \cite{phi2005}.

\begin{figure}[h]
 \includegraphics[height=7cm,width=7cm]{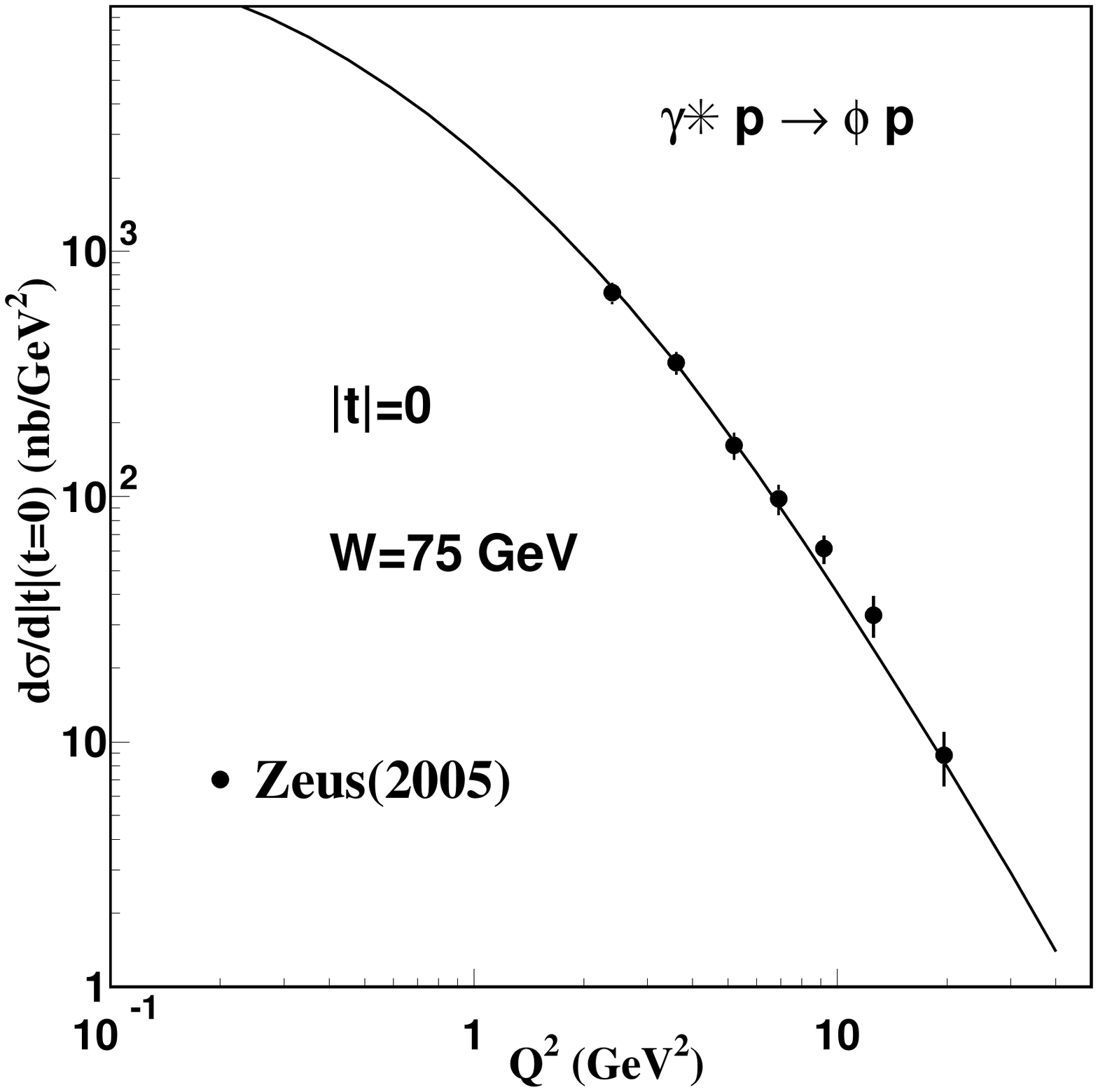}
 \includegraphics[height=7cm,width=7cm]{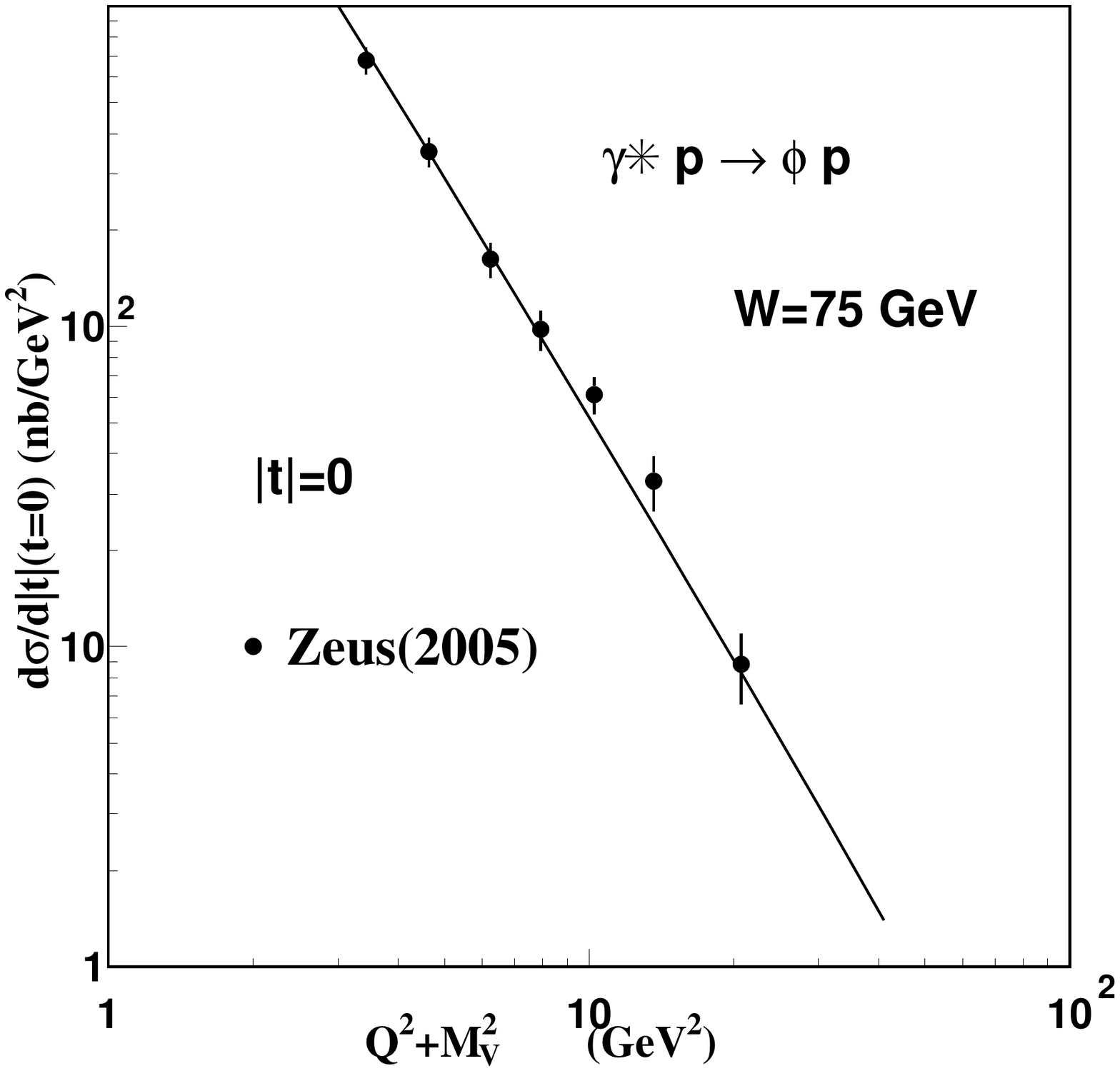}
 \caption{\label{phiforward} $Q^2$   dependence of the
differential cross sections of $\phi$ elastic  in the forward
direction. The solid lines represent our calculation. The second
plot contains the same information, showing the typical behaviour
of a straight line in the variable $Q^2+M^2_V$. }
\end{figure}

There are no published measurements of $d\sigma/d|t|$ in $\phi$
production for nonzero $Q^2$. The Zeus photoproduction data at 94
GeV \cite{phi96,EPJC14}  are shown in Fig. \ref{t_phi}. Our
theoretical calculations give a very satisfactory description.

 \begin{figure}[h]
 \vskip 2mm
 \includegraphics[height=8cm,width=8cm]{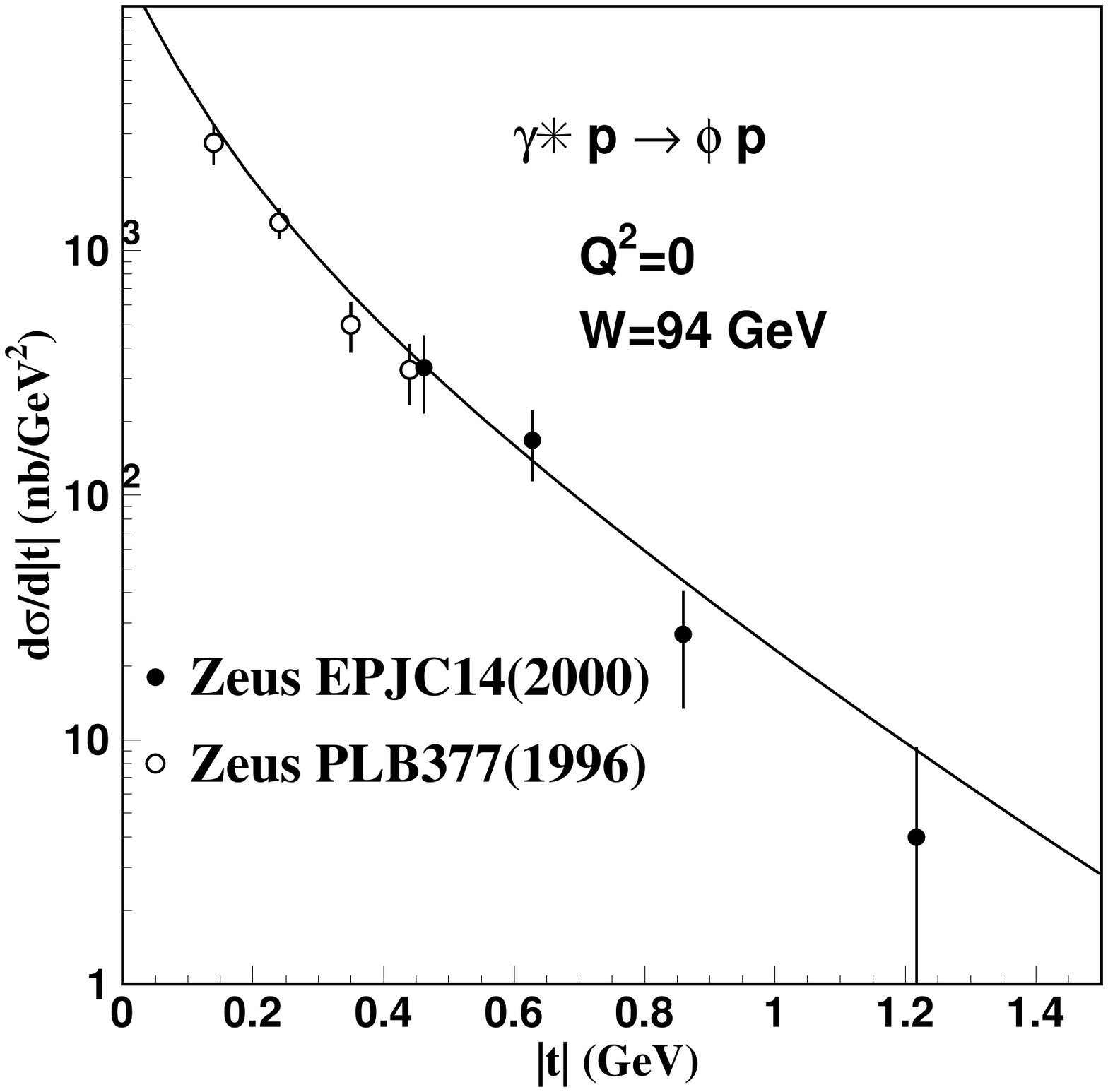}
  \caption{\label{t_phi} $t$ dependence of $\phi$ photoproduction
 cross sections and its theoretical  description in our
nonperturbative  calculation with the stochastic vacuum model.
The published data are from  the Zeus collaboration, at the
energy 94 GeV, first at low t \cite{phi96}
 and then at larger t \cite{EPJC14}.}
 \end{figure}

\subsection{Results concerning several vector mesons }

The quantitative predictions made in a unique way for different
kinds of vector mesons cover several scales of magnitudes in cross
sections. This global coverage is exhibited in Fig.
\ref{allcross}. The same global description is given for the
forward differential cross section shown in Fig. \ref{shiftzero}.
In this figure the charge factors squared $ \hat e_V^2 $, see
Eq. (\ref{e_V}),
for each kind of meson are extracted, making the quantities almost
universal in a $Q^2+M_V^2$ plot. However, the universality is only
approximate, and  our calculation predicts correctly the observed
displacements.

\begin{figure}[h]
 \vskip 2mm
 \includegraphics[height=8cm,width=8cm]{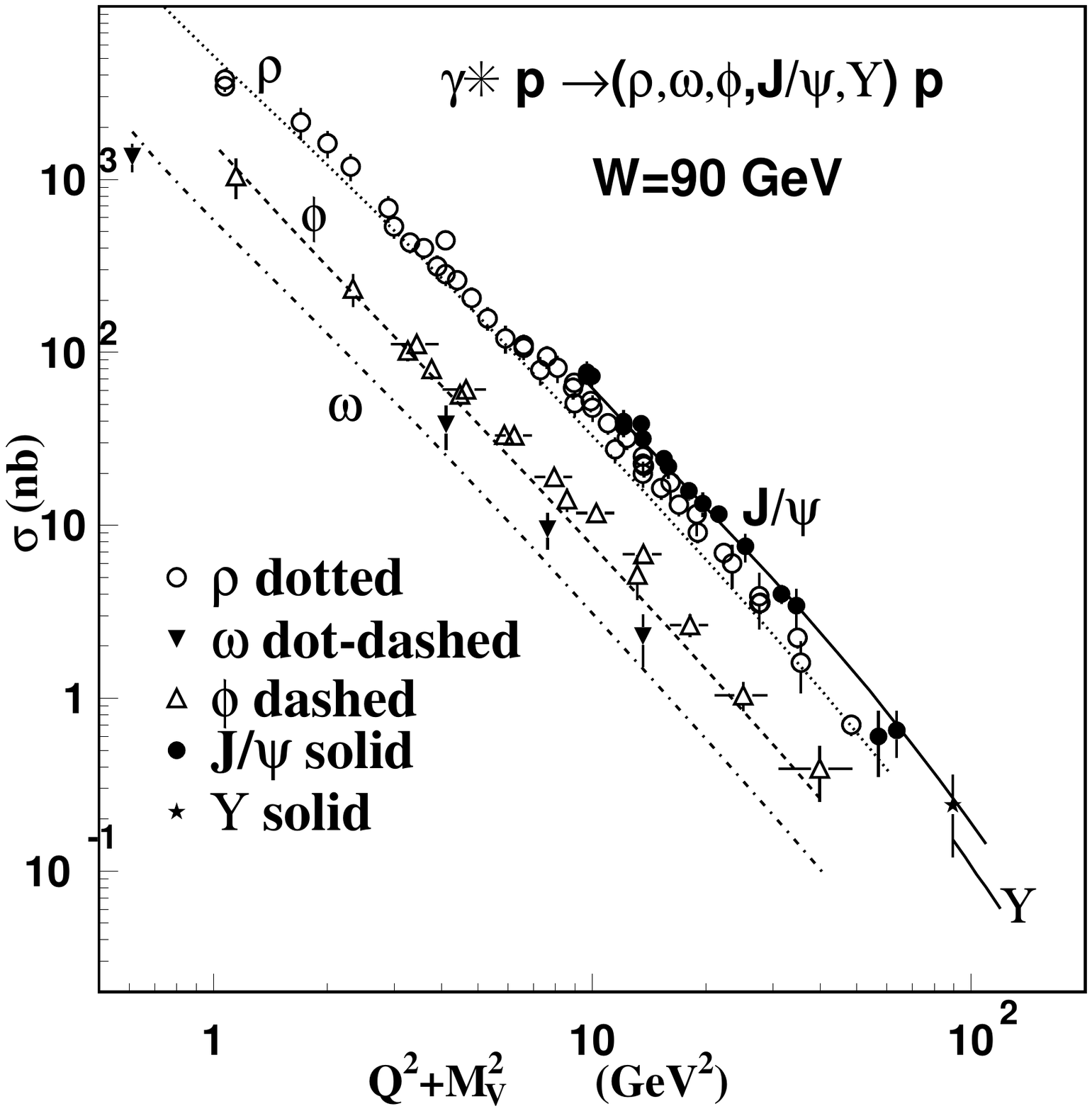}
  \caption{\label{allcross} Integrated elastic cross sections for all
vector mesons at W=90 GeV, as functions of $Q^2+M_V^2$. The lines
represent our theoretical calculations. }
 \end{figure}
\begin{figure}[h]
 \vskip 2mm
 \includegraphics[height=8cm,width=8cm]{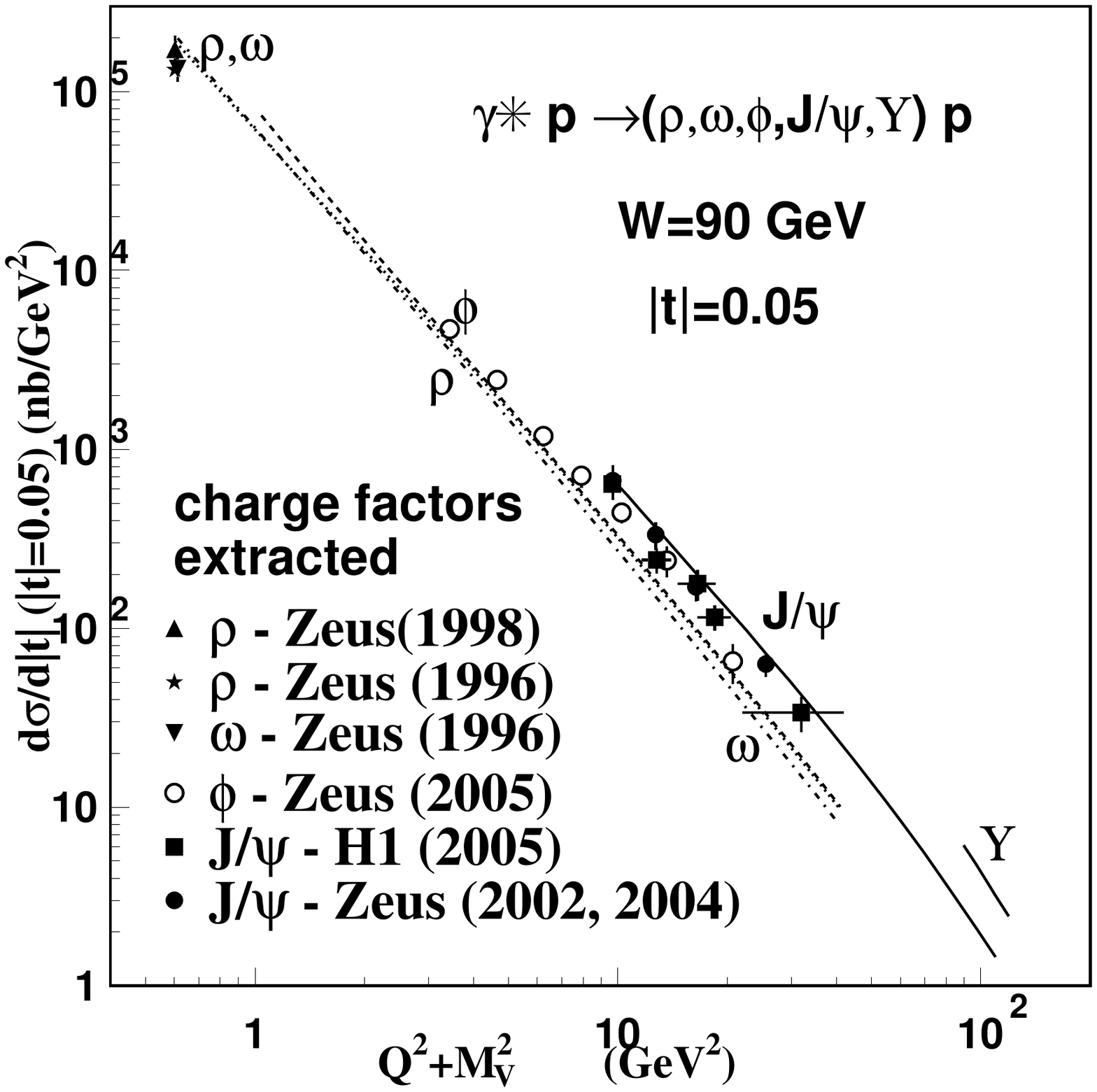}
  \caption{\label{shiftzero} Differential cross sections in a
forward direction for all vector mesons at W=90 GeV, as functions
of $Q^2+M_V^2$, with extraction of charge factors . The lines
represent the theoretical calculations with the stochastic vacuum model
as in Fig. \ref{allcross}. }
 \end{figure}

 The data on the energy dependence of the integrated
cross sections for $\rho, \phi$  and $\psi$  mesons  have been
presented and compared to the theoretical predictions for each case.
The parameter $\delta(Q^2)$ of the suggested simple energy dependence
   $$\sigma(Q^2)={\rm Const.}\times W^{\delta(Q^2)} $$
has also been given in each case. This parametrization is an
approximation valid in a limited energy range, since the true
energy dependence in our calculation is  determined by the
two-pomeron scheme, but it is considered useful in practice.

 We then evaluate $\delta$ using the energy range W=20 - 100
GeV, for all values of $Q^2$. The results are put together in
Fig. \ref{deltas}. We note that all  curves start at the minimum
value $4\times 0.08$ at the same
unphysical point $Q^2+M^2_V=0$ and all are asymptotic to
$\delta=4 \times 0.42$
as $Q^2$ increases. We then have the form of parametrization
\begin{equation}
\label{deltaparam} \delta(Q^2)=0.32+1.36 \frac
{(1+Q^2/M_V^2)^n}{A+(1+Q^2/M_V^2)^n}  \end{equation} and the
values for $A$  and $n$  are given in table \ref{deltatab}.
\begin{table}
\begin{center}
\begin{tabular}{ccccc}\hline
Meson& $\rho$ & $\phi$  & $J/\psi$ & $\Upsilon$     \\
\hline
 $ A$ &   124.926 &  51.9183 &  2.0624 &  0.9307  \\
 $ n $ &  1.239    &  1.239    & 1.12 &   0.22    \\
  \hline
 \end{tabular}
 \end{center}
  \caption{Values of $A$ and $n$ in the exponent $\delta(Q^2)$ from Eq.
 (\ref{deltaparam})} \label{deltatab}
 \end{table}
\begin{figure}[h]
 \vskip 2mm
 \includegraphics[height=8cm,width=8cm]{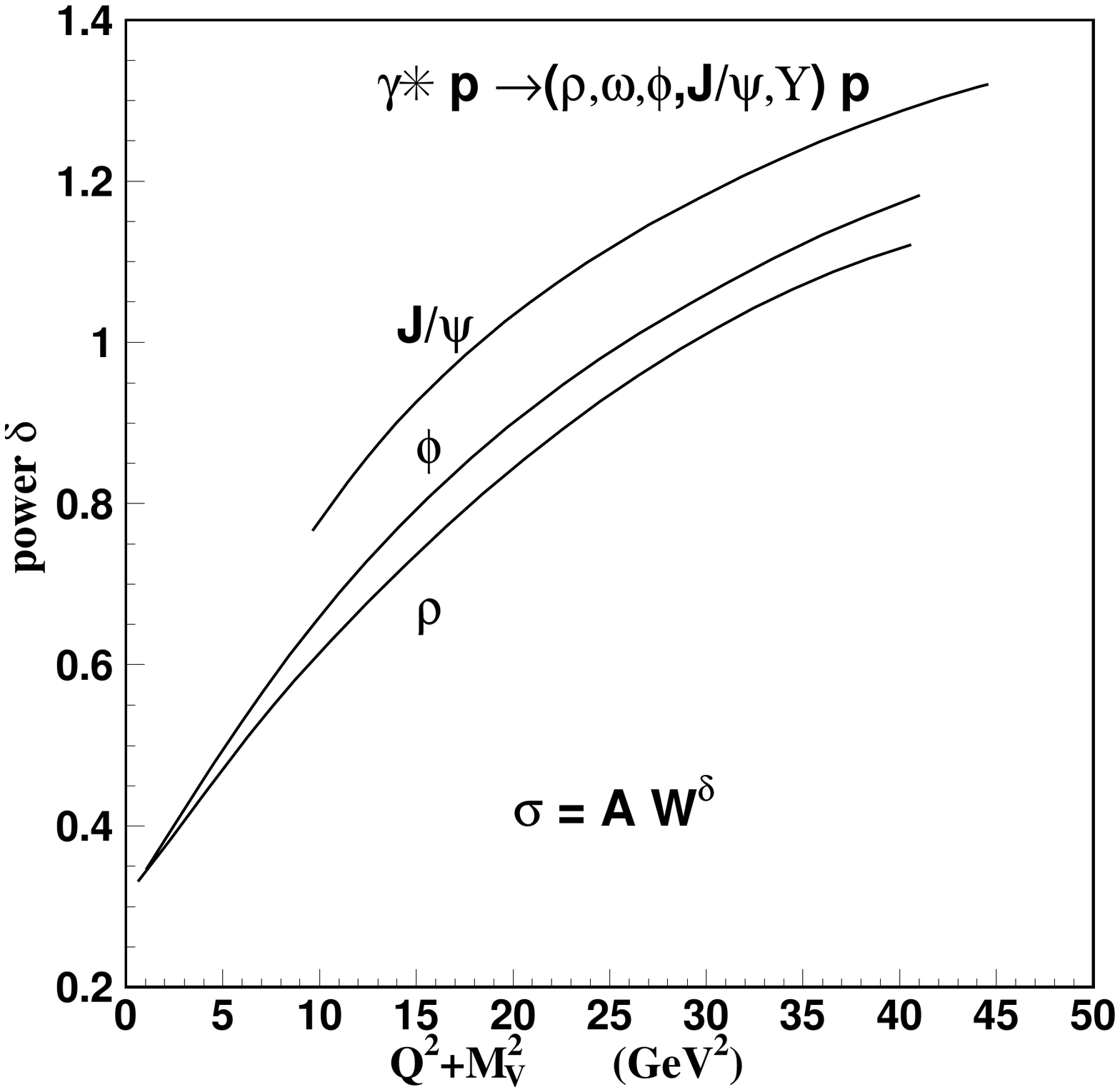}
  \caption{\label{deltas} Parameter $\delta (Q^2)$ of the
energy dependence of the cross sections. }
 \end{figure}

Our model has definite predictions for the $t$ dependence in
differential cross sections. The shape can be conveniently
represented by the form given in Eq.(\ref{ff_eq}).
Fig. \ref{ff_fig} shows the form factor $F(|t|)$ for all vector
 mesons at fixed energy W=90 GeV.

The curvature in the log graph is a prediction of
our framework. The values of the parameters  as functions of $Q^2$
are shown in one of the plots. In Table \ref{t-parameters} some
numerical values of  $ a, b$  are given.

 \begin{figure}[h]
 \vskip 2mm
 \includegraphics[height=7cm,width=7cm]{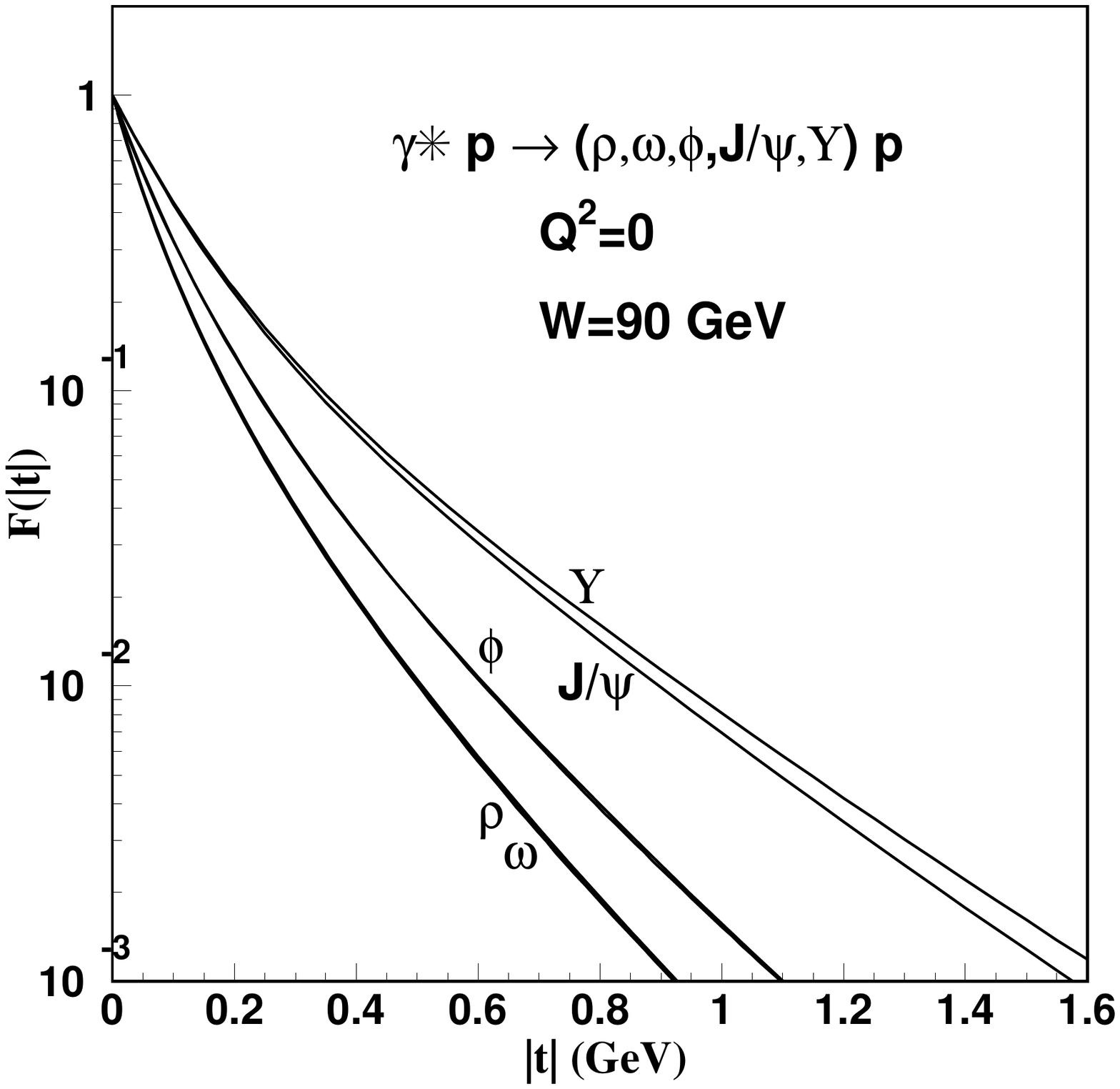}
 \includegraphics[height=7cm,width=7cm]{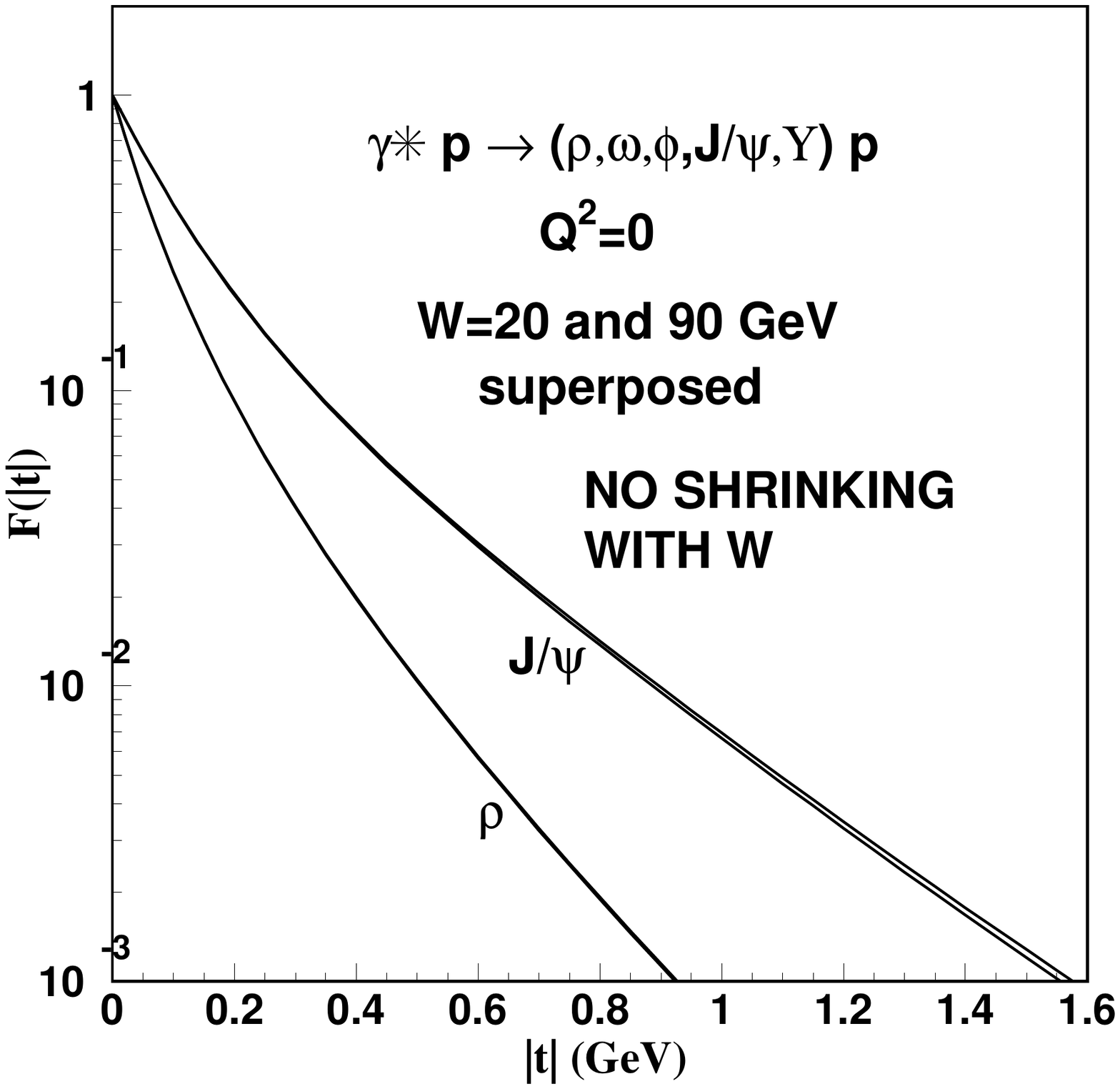}
 \caption{\label{ff_fig} Form factor $F(|t|)$ of the $|t|$
distribution in the elastic differential cross sections photoproduction
of vector mesons, with the characteristic
curvatures in the log plot predicted in our calculations.
 The curves follow the shapes given by Eq.(\ref{ff_eq}),
with $a(Q^2)$ and $b(Q^2)$ represented in the second figure.
Specific numerical values of the parameters for $Q^2=0~ , ~
 10 ~ {\rm and} ~  20 ~ {\rm GeV}^2$  are given in
Table \ref{t-parameters}. The values in the limit of very
large $Q^2$ are $a=4.02~{\rm GeV}^{-2}$ and
$ b=1.60 ~{\rm GeV}^{-2}$.  }
\end{figure}

\begin{table} [h]
 \caption{ \label{t-parameters} Some values
(in ${\rm GeV}^{-2}$)  of the parameters
$a $ and $b$ of $\frac{d\sigma}{d|t|}=
\Big[\frac{d\sigma}{d|t|}\Big]_{t=0} \times
\frac{e^{-b|t|}}{(1+a|t|)^2}$~ in Eq.(\ref{ff_eq}). The full 
$Q^2$ dependence is  shown in Fig. \ref{ff_fig}.  }
 \begin{center}
 \begin{tabular}{l c c c c c c  } \hline
 \multicolumn{1}{c}{ }
          &\multicolumn{2}{c} { $Q^2 = 0$ } &
           \multicolumn{2}{c} { $Q^2=10 ~ {\rm GeV}^2$ } &
           \multicolumn{2}{c} { $Q^2=20 ~ {\rm GeV}^2$ }\\
\hline
  Meson &$a $&$b $& $a $& $b $ & $a $& $b $  \\
 \hline
 $\rho(770)$   & 7.06   &  3.09   & 4.349  & 1.996 & 4.145 & 1.846 \\
 $\omega(782)$ & 7.20   &  3.08   & 4.323  & 1.990 & 4.141 & 1.835 \\
 $\phi(1020)$  & 5.40   &  2.77   & 4.211  & 1.934 & 4.091 & 1.807 \\
 $J/\psi(1S)$  & 4.06   &  1.75   & 4.026  & 1.696 & 4.022 & 1.670 \\
 $\Upsilon(1S)$& 4.03   &  1.61   & 4.027  & 1.606 & 4.025 & 1.604 \\
 \hline
 \end{tabular}
 \end{center}
 \end{table}

As  the virtuality $Q^2$ grows, the ranges of the overlap functions
decrease, and the electroproduction cross sections of all mesons
become flatter, all tending together to the shape characteristic
of the $\Upsilon$ meson, with same limiting values
$a=4.02 ~ {\rm GeV}^2  $ and $ b=1.60 ~ {\rm GeV}^2 $ for the parameters.
The limiting shape  of the distribution for very large $Q^2$ is
illustrated in  Fig. \ref{ff100fig}, where we see all vector mesons
superposed. In the figure we draw the bit of straigth line
representing the slope considered as the average for the interval
from $ |t|=0 $ to $|t|=0.2 ~ {\rm GeV}^2 $.
As indicated inside the plots, the calculations of
form factors presented in the figures are made for $W=90$ GeV.
In the second plot presented in Fig. \ref{ff100fig} we show the
(absence of) dependence of the form factor on the energy W.
Thus we predict that there is no shrinking of the forward peak
in $d\sigma/d|t|$ as the energy increases. The experimental
data are not yet sufficient to test all these predictions.

 \begin{figure}[h]
 \vskip 2mm
 \includegraphics[height=7cm,width=7cm]{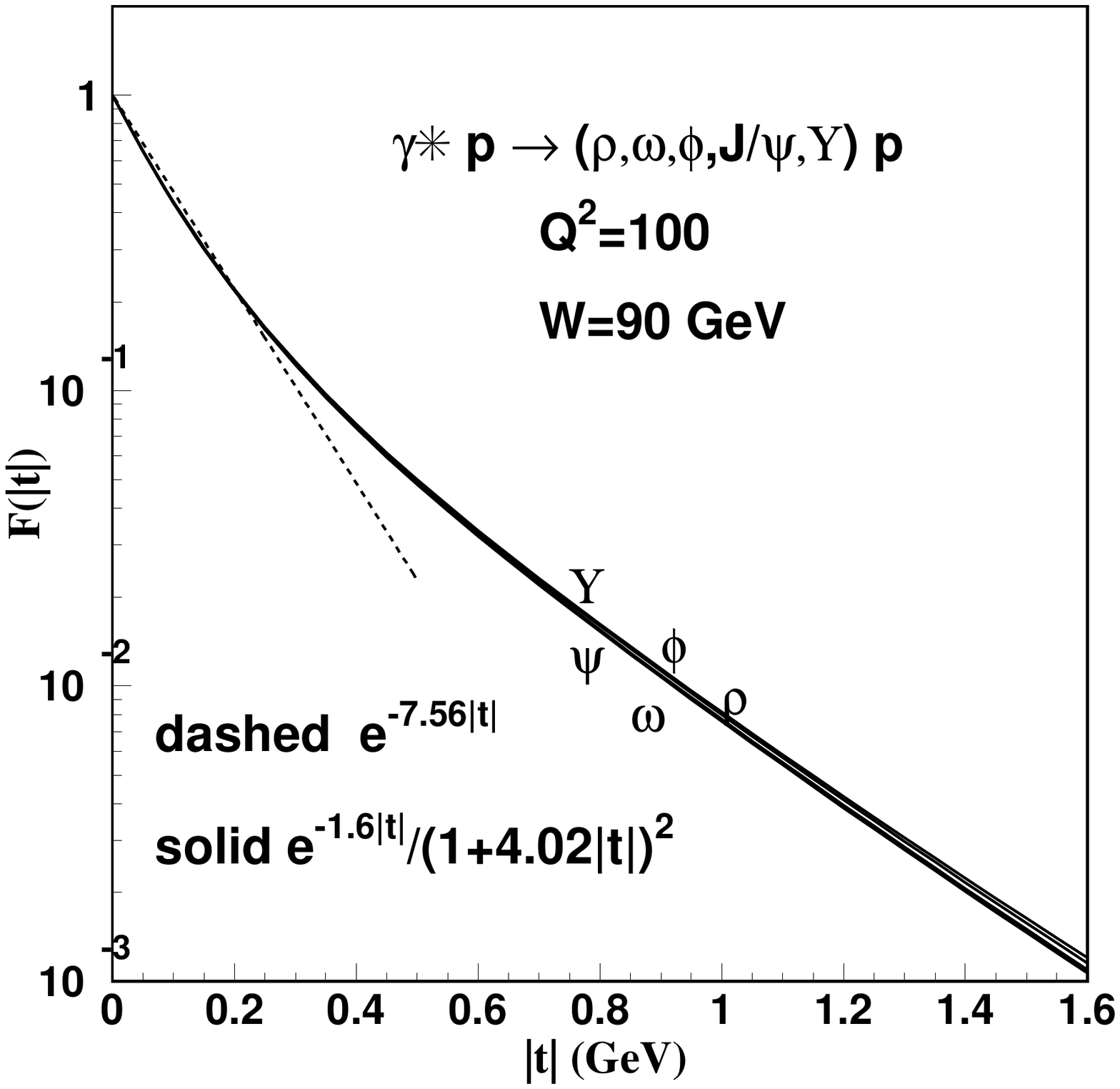}
 \includegraphics[height=7cm,width=7cm]{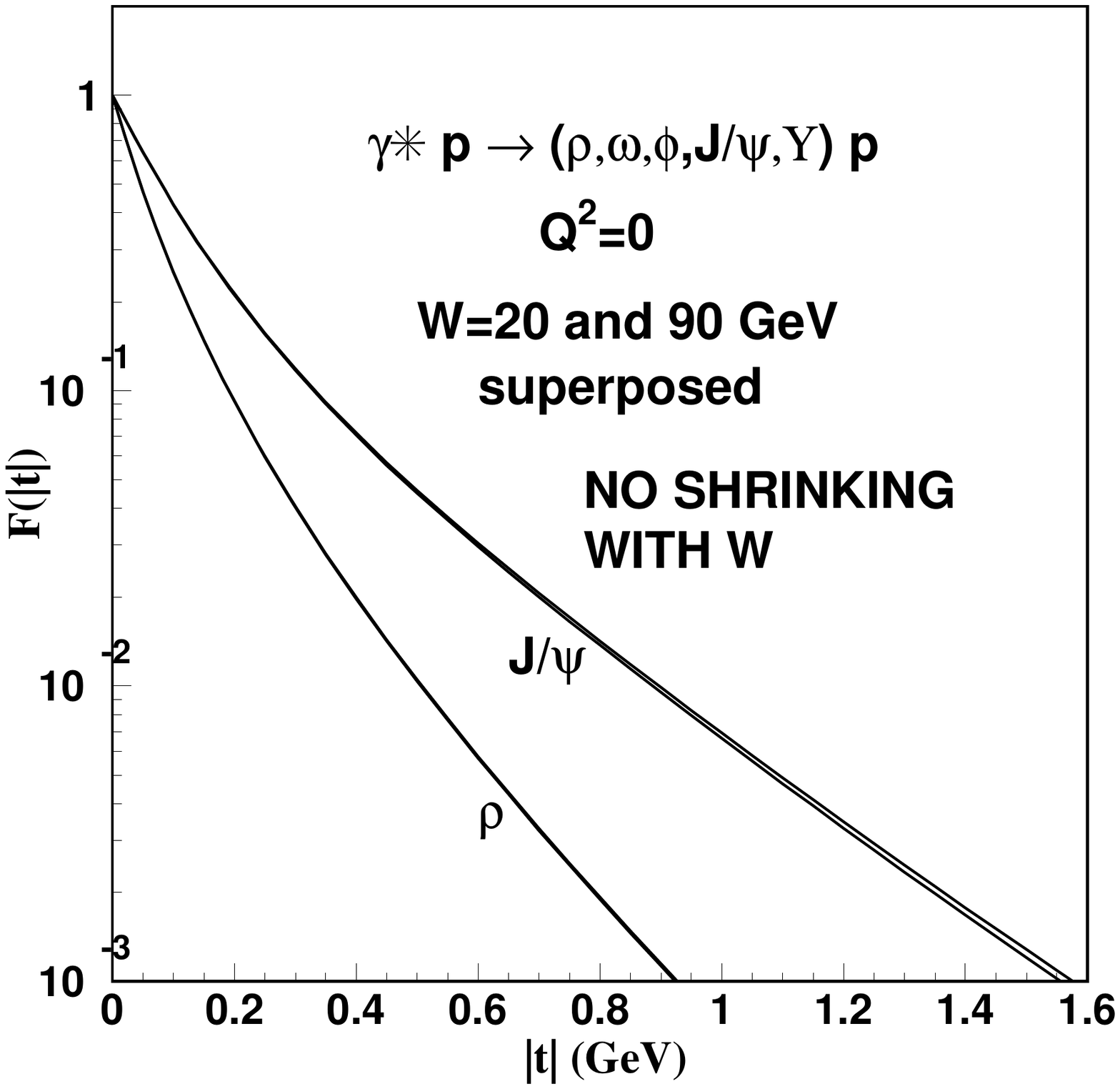}
 \caption{\label{ff100fig}
The plot in the left hand side shows the shape of the $|t|$
distribution in the differential cross section for
electroproduction common to all vector mesons  for very large $Q^2$.
The straight line passes through the points $|t|$ equal to
zero and 0.2 GeV$^2$, indicating the average slope that would
be measured in this limit. The plot in the right hand side
shows that $F(|t|)$ does not depend significantly on the
energy, and thus measurement of slopes at fixed $Q^2$
would give a constant value for each vector meson.}
\end{figure}

\section{Relation with other approaches } \label{other}  

Our work runs parallel to other developments , of  mainly nonperturbative 
 nature,  based on the colour  dipole dynamics.  These  treatments, 
which started with studies of  $\gamma^* p$ total cross section \cite{GBW99a}
and soon were extended to diffractive deep inelastic scattering \cite{GBW99b},
are based on the  assumption that the photon fluctuates into a 
$q\bar q$ pair, which then scatters with the proton, according to a  
properly built dipole cross section $\sigma_{\rm dipole}$.  The $q\bar q$ 
dipole then recombines to form  again a  photon, or eventually a vector 
meson, in the final state. Photons, either real or virtual, enter in a well 
defined way through their QED  wave functions, and the dipole cross section 
is built with the necessary  ingredients to describe the observed 
phenomenology and, as much as posible,  to follow QCD prescriptions.  
The proton structure  is introduced to give an impact parameter dependence 
that allows description of $t$ dependence in a predictable way.

Concerns with observed saturation effects in the $Q^2$  and $x$ dependences 
led to an ansatz for $\sigma_{\rm dipole}$  of the form
\begin{equation}
\sigma_{\rm dipole}(x,r)=\sigma_{0} \big(1-\exp{(-r^2 Q_s^2(x)/4)} \big) ~, 
\end{equation} 
where $\sigma_0$ is a constant and $Q_s(x)$ denotes $x$ dependent 
saturation scale. For processes governed  by quarks of  mass $m_f$ , the
Bjorken scale $x$  is taken as 
\begin{equation} 
x_{m_f}= \frac{Q^2}{Q^2+W^2} \big( 1+\frac{4 m_f^2}{Q^2}  \big)  ~ .
\end{equation}

This kind of dipole cross section worked  well for what it was aimed for, 
giving  a  description  of the HERA data  on inclusive (related to $F_2$)  
and diffractive $\gamma^* p$ processes. In an effort to improve the model 
including  scaling violations  and QCD  evolution, Bartels, Golec-Biernat 
and Kowalski \cite{BGBK02}
replaced the saturation scale $Q_s^2$  by a gluon density obeying DGLAP 
evolution. The interplay between saturation and  evolution improved fits 
of the $F_2$  data , especially for large $Q^2$.

Another improvement  of the original   saturation model came with the 
Color Glass Condensate model \cite{CGC04}, 
dealing  with the question of the low-$x$ behaviour 
near the saturation region through BFKL dynamics.  These calculations 
were successful as an extension of the  model, although they did not 
include charm contributions.  

An important line of development of the phenomenological color dipole 
models came with the introduction of effects of the proton shape, with 
explicit impact parameter dependence in the dipole scattering, by 
Kowalski and Teaney \cite{KT03},  
and consequently with possibility of treatment  of $t$ distributions. 

This work was extended by Kowalski, Motyka and Watt  to the description 
of photo- and electroproduction of  vector mesons \cite{KMW06}, going 
beyond $F_2$ and DVCS. The authors use the same kind of dipole 
cross section as KT , based on Glauber-Mueller:
\begin{equation}
\frac{d \sigma_{\rm dipole}}{d^2\vec{b}}=2\big[1-
\exp{\bigg(- \frac{\pi^2}{2 N_c} r^2 \alpha_s(\mu^2) x g(x,\mu^2) T(b)\bigg)}\big] ~ . 
\end{equation}
The scale $\mu^2$ is related to the dipole size, and the gluon density is 
evolved from $\mu_0^2$ to $\mu^2$ using LO DGLAP evolution without quarks. 
Application of this model requires a number of ingredients and parameters, 
as form of  gluon density at $\mu_0$, effective $x$ for heavy quarks, proton 
shape $T(b)$, b-dependence in the form of $\sigma_{\rm dipole}$. The 
paper describes part of the recent HERA data on   photo- and  
electroproduction of vector mesons that we discuss in the present paper, 
with interesting similarity of results. 

A fundamental and original approach for the implementation of the dipole 
treatment of vector meson production, which provides the framework for the 
present work, was developed by the Heidelberg group \cite{DGKP97,DGP98,KDP99}.
This approach calculates the basic loop-loop interaction using the 
stochastic vacuum model \cite{Dos87,DS88,DFK94}, which is a genuine 
nonperturbative treatment that allows connections with other branches of 
nonperturbative QCD, especially with the fundamental and striking feature 
of confinement. The method is based on functional methods \cite{Nac91},
and has been used recently for investigations on the foundations and 
limitations of the dipole model \cite{EN05,EN06a,EN06b}.  
 
The same approach using the loop-loop amplitude and the stochastic 
vacuum model was used consistently by Donnachie and Dosch for calculations 
of DVCS \cite{DD01} and   structure functions \cite{DD02} . 
  
The treatment of photoproduction of all vector mesons through a dipole 
Pomeron model respecting the unitarity bounds by Martynov, Predazzi and 
Prokudin \cite{MPP03} nicely described $ W$ and $t$ dependences in elastic 
photoproduction  processes, and was soon extended to electroproduction 
\cite{FJPP03}. 
 
A separation of hard (small)  and soft (large)  dipole interactions, 
testing several prescriptions for the dipole cross section and several 
forms of wave function, was used  by  Forshaw, Sandapen and Shaw 
\cite{FSS04,FSS06} to describe  energy and $Q^2$ dependences  of  the 
 total $\gamma^*p$ cross section, diffractive electroproduction, DVCS and  
exclusive $J/\psi$ electroproduction.  The separation of dipoles in two 
classes is similar to the two-pomeron model  adopted in the present work. 

As a whole, these nonperturbative models for $\gamma^*$ induced processes,
all based in the dipole picture, are phenomenologically satisfactory. 
This is not surprising, since it has already been shown \cite{EFVL} 
that many features of these processes, particularly the $Q^2$ 
dependence, are reproduced by the overlap of the light cone wave functions 
of photons and mesons folded with the basic $r^2$ behaviour of the dipole
cross sections. 

Unfortunately the presently available data are not sufficient to 
discriminate details of different approaches, particularly in the 
incorporation of energy dependence.

\section{Summary and discussion}

We have shown the predictions for elastic electroproduction processes using   two basic   ingredients: \\
1) the overlaps of photon and meson wave functions, written 
as packets of quark-antiquark dipoles, with protons 
described also as packets of dipoles (in a convenient 
diquark model for the nucleon), \\
2) the interaction of two dipoles described in terms of geometric variables in an impact parameter representation 
of the amplitudes
based on nonperturbative properties of the QCD gluon field. 

These quantities put together and integrated over the distribution of dipoles
in initial and final states lead to a correct description of
the data concerning all vector mesons.

In all cases the variations with energy are very well described by the Regge picture, with soft and hard pomerons coupled to large and small dipoles, respectively. We recall that in approaches mainly based on perturbation theory the energy (or $x$) dependence
is introduced through the gluon distribution in the proton.

Each of the different mesons enter in the calculation
characterized only by the masses and charges of its quark
contents, and with their normalized wave function individualized
only by the corresponding electromagnetic decay rate (related to
the value of the wave function at the origin).

The specific nonperturbative input is the stochastic vacuum model,
which has been  successfully applied
in many fields, from hadron spectroscopy to high energy scattering.
The basic interaction of two dipoles depends only on universal
features of the QCD field, which are the numerical values of the
gluon condensate and of the correlation length of the finite range
correlations. These two quantities have been determined by lattice
investigations and tested independently in several instances of
phenomenological use  of the dipole-dipole interaction.

As has already been pointed out \cite{EFVL}, the main features of
the  $Q^2$ dependence of electroproduction of vector mesons are
contained in the overlap integral. For values of $Q^2$ attainable
at present this overlap integral is determined by perturbative
QCD, through the photon wave function, and by nonperturbative QCD,
through the meson wave function. The production of transversely
polarized mesons is determined by the meson wave function even for
very high values of $Q^2$. The importance of nonperturbative effects
in vector meson production has also been stressed in \cite{DGS01},
where the perturbative two gluon exchange has been supplemented
by an exchange of nonperturbative gluons. In our approach, which
incorporates both perturbative and nonperturbative effects, we are
able to give a fair overall description of all observables of vector
meson production: the energy dependence, the $Q^2$ dependence, the
ratio of longitudinal to transverse mesons and the angular distribution.

In this paper our calculations are compared to the large amount of HERA data for 
$\rho,\omega,\phi,{\rm J}/\psi $ and $\Upsilon$ mesons.

\begin{acknowledgement} Both authors wish to thank DAAD (Germany), 
CNPq (Brazil) and FAPERJ (Brazil) for support of the scientific 
collaboration program between Heidelberg and Rio de Janeiro groups
working on hadronic physics. One of the authors (EF) is grateful to 
CNPq (Brazil) for research fellowship and grant. 
\end{acknowledgement}

\appendix 

\section{QCD and wave function parameters}

Here we recall some expressions and the numerical values 
of quantities  used in the nonperturbative contributions,  
which have been unchanged for many applications. 
 
The  loop-loop scattering amplitude 
$ S(b,W,z_1,\vec R_1,z_2=1/2,\vec R_2)$
determining the essential  quantity $ J(\vec q,W,z_1,\vec R_1)$ in Eq. (\ref{int2}),
can be calculated using the stochastic vacuum model. The essential input 
parameters of this model are the correlation length $a$ of the gauge invariant 
two gluon correlator and the gluon condensate $\langle g_s^2FF\rangle$. 
 These quantities  have been determined in lattice calculations \cite{DGM99}. 
The numerical values used in this and previous papers are
\begin{equation}
a=0.346~{\rm fm}~~~~~~~~~~~\langle g_s^2FF\rangle\, a^4=23.5 ~.
\end{equation}
The constant $\kappa=0.74$ appearing in the correlation functions preserves 
its value determined by lattice calculations.

The energy dependence is based on a two-pomeron model, small dipoles with 
size $R<R_c$ couple to the hard pomeron with an intercept 
$\alpha_{P\,h}(0) = 1.42$, whereas
large dipoles with $R>R_c$ couple to the soft pomeron with an intercept  $\alpha_{P\,s}(0) = 1.08$. The transition radius $R_c$ has been determined 
from the $x$-dependence of the proton structure function to $R_c=0.22 $ fm. 

For the proton wave function occurring in Eq. (\ref{int2}) we use a 
diquark-quark Gaussian wave function  with the transverse radius 
$R_P=0.75$ fm.

The light quark masses which can simulate confinement effects in the otherwise perturbative photon wave function \cite{DGP98} are
\beq
m_u=m_d=0.2~{\rm GeV} ~~~~~~~~~~~~~~m_s=0.3 ~{\rm GeV} ~ . 
\enq
For the heavy quarks we take the renormalized masses in the $\overline{MS}$ scheme
\beq
m_c=1.25 ~ {\rm GeV} ~~~~~~~~~~~~~~m_b=4.2 ~ {\rm GeV} 
\enq

The full form  of photon and vector meson wave functions used in our work,
 including the helicity dependences, have been presented before 
\cite{DF02,EFVL}, with two forms for the vector mesons: 
the Bauer-Stech-Wirbel (BSW) \cite{BSW87}, and the 
Brodsky and Lepage (BL) \cite{BL80} wave functions, where the 
 separate $r,z $ dependences are respectively 
  \begin{eqnarray} \label{BSW}
&& \phi_{BSW}(z,r) =      \\
&&\frac{N}{\sqrt{4 \pi~}} ~\sqrt{z \bar{z}~}~
      \exp\Big[-\frac{M_V^2(z-\frac{1}{2})^2}{2 \om^2}\Big]
     \exp[-\frac{\om^2 r^2}{2}] ~ ,  \nonumber
 \end{eqnarray}
(here $M_V$ represents the vector meson mass) and 
  \beqa \label{BL}
&&  \phi_{BL}(z,r) =    \\
&& \frac{N}{\sqrt{4 \pi~}}
 \exp\Big[-\frac{m_f^2(z-\frac{1}{2})^2}{2z\bar{z}\om^2 }\Big]\exp[-2 z\bar{z}\om^2 r^2]~ ,  \nonumber
 \enqa
with  $m_f$ representing the quark mass. 

The parameters N (normalization) and $\omega$ (that fixes the extension) 
are determined  using the  electromagnetic decay rates of the vector 
mesons. Their values are collected \cite{EFVL} in  Table \ref{WFparam:1}.
  \begin{table*} [t] 
  \caption{ \label{WFparam:1}  Parameters of the vector meson wave functions }
 \begin{center}
 \begin{tabular}{|l|c|c c c c|c c c c|}\hline
\multicolumn{1}{|l|}{} & \multicolumn{1}{|c|}{} &\multicolumn{4}{c |}{\bf BSW} &
           \multicolumn{4}{c|}{\bf BL}  \\
\multicolumn{1}{|l|}{}& \multicolumn{1}{|c|}{} 
          &\multicolumn{2}{c}{\bf transverse} &
           \multicolumn{2}{c|}{\bf longitudinal}
          &\multicolumn{2}{c}{\bf transverse} &
           \multicolumn{2}{c|}{\bf longitudinal} \\
   Meson     & $f_V$ (GeV) &$~ \om$(GeV)& $N$    &$~ \om$(GeV)& $N$
                        &$~ \om$(GeV)& $N$    &$~ \om$(GeV)& $N$   \\  \hline
 $\rho(770)  $& $0.15346$ &$0.2159 $ &$5.2082$&$0.3318   $    &$4.4794$
                       &$0.2778 $  &$2.0766$&$0.3434$   &$1.8399$ \\
 $\omega(782)$& $0.04588$ & $0.2084 $ &$5.1770$&$0.3033   $    &$4.5451$
                       &$0.2618 $  &$2.0469$&$0.3088$   &$1.8605$ \\
 $\phi(1020) $& $0.07592$ & $0.2568 $ &$4.6315$&$0.3549   $    &$4.6153$
                       &$0.3113 $  &$1.9189$&$0.3642$   &$1.9201$ \\
 $J/\psi(1S) $&$ 0.26714 $ & $0.5770 $ &$3.1574$&$0.6759   $    &$5.1395$
                       &$0.6299 $  &$1.4599$&$0.6980$   &$2.3002$ \\
 $\Upsilon(1S)$& $ 0.23607$ & $1.2850 $ &$2.4821$&$1.3582   $    &$5.9416$
                       &$1.3250 $  &$1.1781$&$1.3742$   &$2.7779$ \\
 \hline  \end{tabular}  \end{center} 
 \end{table*} 									 

After summation over helicity indices, the overlaps of the
photon and vector meson wave functions
\beq
 \rho_{\ga^* V,\lambda} (Q^2;z_1,{\mathbf R}_1) =
  \psi_{V\lambda}(z_1,{\mathbf R}_1)^*\psi_{\ga^* \lambda}(Q^2;z_1,{\mathbf R}_1)
\label{over}
\enq
that appear in Eq.(\ref{int})   are given by
 \beqa \label{overBSW1}
&& \rho_{\ga^* V,\pm 1;BSW}(Q^2;z,r)    
 = \hat e_V\frac{\sqrt{6\alpha}}{2\pi} \\
&& \Big( \ep_f ~\om^2 r \big[z^2+\bar{z}^2\big] K_1(\ep_f ~r)
     + m_f^2 K_0(\ep_f ~r)\Big)~\phi_{BSW}(z,r)  
\nonumber 
 \enqa
 and
 \beqa \label{overBL1}
 && \rho_{\ga^* V ,\pm1;BL}(Q^2;z,r)=\hat e_V\frac{\sqrt{6\alpha}}{2\pi} \\
&& \Big(4 \ep_f ~\om^2 rz\bar{z} \big[z^2+\bar{z}^2\big] K_1(\ep_f ~r)
    + m_f^2 K_0(\ep_f ~r)\Big)~\phi_{BL}(z,r)  
  \nonumber  \enqa
 for the transverse case, BSW and BL wave functions respectively.
  For the longitudinal case we can write jointly
 \beq \label{overlong}
\rho_{\ga^* V ,0;X}(Q^2;z,r) = -16 \hat e_V\frac{\sqrt{3\alpha}}{2\pi}\om
     z^2 \bar{z}^2~Q~ K_0(\ep_f ~r) \phi_X(z,r) ~ ,
 \enq 
 with $X$   standing for BSW or BL.
The effective quark charges $\hat e_V$ are $1/\sqrt{2}$ for $\rho$,
$1/3\sqrt{2}$ for  $\omega$, $-1/3$ for $\phi$ and $\Upsilon$ and $2/3$ 
for $\psi$.  The quantity
 \beq  \label{epsilon}    \ep_f=\sqrt{z(1-z) Q^2+m_f^2} ~ \enq
and the modified Bessel functions are introduced by the photon 
wave functions.

\end{document}